\documentclass[useAMS,usenatbib]{mn2e}

\usepackage{amssymb}
\usepackage{float}
\usepackage{bm}
\usepackage{epsfig}
\usepackage{soul}
\usepackage{subcaption}
\usepackage{multirow,balance}
\usepackage{amsfonts}
\usepackage{amsmath}
\usepackage{aas_macros}
\usepackage{graphicx}
\usepackage{natbib}
\usepackage{color}
\usepackage{listings}
\usepackage{hyperref}  
\usepackage{cases}
\usepackage{tikz}
\usetikzlibrary{decorations.pathreplacing}

\hypersetup{dvips, colorlinks=true, linkcolor=black, citecolor=black, filecolor=black, urlcolor=black}

\definecolor{grey}{rgb}{0.75,0.75,0.75}
\definecolor{Orange}{rgb}{1.0,0.5,0.15}
\definecolor{brown}{rgb}{0.7,0.25,0.0}
\definecolor{pink}{rgb}{1.0,0.5,0.5}
\definecolor{darkerred}{rgb}{0.8,0,0}
\definecolor{darkerblue}{rgb}{0,0,0.8}
\definecolor{Blue}{rgb}{0,0.08,0.65}
\definecolor{Red}{rgb}{0.65,0.08,0.05}
\definecolor{Green}{rgb}{0.15,0.45,0.25}

\begin{document}

\setcounter{tocdepth}{3}


\title[Secular resonant dressed orbital diffusion]
{  Secular resonant dressed orbital diffusion I :\\
method and WKB limit for tepid discs
}

\author[Jean-Baptiste Fouvry, Christophe Pichon, Simon Prunet]{Jean-Baptiste Fouvry$^{1,2}$, Christophe Pichon$^{1,2,3}$, Simon Prunet$^{1,2,4}$
\vspace*{6pt}\\
\noindent$^{1}$ Institut d'Astrophysique de Paris, CNRS (UMR7095), 98 bis Boulevard Arago, 75014, Paris, France\\
$^{2}$ UPMC Univ. Paris 06, UMR7095, 98 bis Boulevard Arago, 75014, Paris, France\\
$^{3}$ Institute of Astronomy, University of Cambridge, Madingley Road, Cambridge, CB3 0HA, United Kingdom\\
$^{4}$ CFHT Corporation, 65-1238 Mamalahoa Hwy, Kamuela, Hawaii 96743, USA\\
}

\date{\today}
\label{firstpage}
\pagerange{\pageref{firstpage}--\pageref{lastpage}}

\maketitle

\begin{abstract}
The equation describing the  secular diffusion of a self-gravitating collisionless system induced by an exterior perturbation
is derived while assuming that the timescale corresponding to secular evolution is much larger than that corresponding 
to the natural frequencies of the system. 
Its two dimensional formulation for a tepid galactic disc  is also derived using the epicyclic approximation.
Its  WKB limit is found while  assuming that only tightly wound transient spirals are sustained by the disc. It yields a simple quadrature for the diffusion coefficients which provides a straightforward  understanding of the loci of maximal diffusion within the disc. 
\end{abstract}


\section{Introduction}
\label{sec:introduction}

Understanding the secular dynamical evolution of galaxies over cosmic time has been a long standing subject of interest. 
Indeed, self gravitating collisionless systems such as galaxies  may, over cosmic times, change their kinematical structure as they respond 
secularly to their evolving environment, in a manner which depends both on their internal orbital structure, but also on how this structure 
resonates with  its environment or with itself.  
It is therefore critical
to distinguish in the  physical properties of galaxies the contributions from the cosmic environment (nurture) and its induced pertubations from the ones coming from the intrinsic properties of the galaxies (nature). 
For thermodynamically improbable cold systems such as galactic discs, their gravitational susceptibility should also play a specific role which 
must be taken into account when studying their long term evolution. 

To tackle this question, one can  rely on numerical N-body simulations of higher resolutions to take into account non-linear physical processes \citep[e.g.][]{Dubois2014} or perform idealized well-crafted numerical experiments \citep{SellwoodAthanassoula1986,Earn1995,Sellwood2012}. With such statistical investigations, one can assess the importance and the role of the orbital structure of a galactic disc to drive its secular evolution. 
Angle-action variables \citep{born1960mechanics,Goldstein,BinneyTremaine2008} and the matrix method \citep{Kalnajs2} also allow us to  take into account   the self-gravitating amplification of such collisionless systems.
 From this analytical framework, one should therefore be able to derive  flexible qualitative and quantitative equations describing the secular dynamics of discs, without relying on the implementation of the corresponding demanding numerical models. 


This topic of secular  evolution has  been addressed via the dressed Fokker-Planck equation, where the source of secular evolution for a self-gravitating system is taken to be potential fluctuations from an external bath, e.g. corresponding to the cosmic environment. \cite{Binney1988} computed the first and second-order diffusion coefficients describing the orbits deviation induced by fluctuations in the gravitational potential. \cite{Weinberg93} showed the importance of self-gravity on the nonlocal and collective relaxation of stellar systems. \cite{Weinberg2001a} and \cite{Weinberg2001b} considered the dressed gravitational amplification of Poisson shot noise in stellar systems and the impact of the properties of the noise processes. \cite{MaBertschinger2004} used a quasi-linear approach to investigate dark matter diffusion induced by cosmological fluctuations. \cite{Pichon2006} sketched a time-decoupling approach to solve the collisionless Boltzmann equation and applied it to the statistical study of dynamical flows through dark matter haloes. \cite{Chavanis2012EPJ} considered the evolution of homogeneous collisionless systems forced by an external perturbation. \cite{Nardini2012} also considered the evolution of such long-range interacting systems when driven by external stochastic forces.

Using an argument based on timescale decoupling inspired from \cite{Pichon2006}, we present here a careful and detailed derivation of the secular resonant dressed orbital diffusion equation for a general collisionless self-gravitating system undergoing external perturbations \footnote{We also recover a missing ${ 1/2 }$ factor absent from this previous work, due to an error in temporal integration bounds.}. We then develop this formalism for the secular evolution of an infinitely thin galactic disc. In order  to  circumvent the complex direct or analytical calculation of the modes of a galactic disc carried only for a small number of disc models \citep{Zang1976,Kalnajs1977,Goodman1988,Weinberg1991,Vauterin1996,Pichon1997,Evans1998,Jalali2005}, we then  rely on the WKB approximation (\citealt{WKB}; \citealt{Toomre1964}; \citealt{Kalnajs1965}; \citealt{Lin1966}) to obtain a tractable algebraic expression for both the gravitational susceptibility of the system and the associated diffusion coefficients.  Within the realm of this approximation, which should apply to 
cold enough discs,
the diffusion coefficient reduce to simple quadratures.
  
The paper is organized as follows. Section~\ref{sec:seculardiffusionequation} presents a  derivation of the general dressed Fokker-Planck equation for a perturbed self-gravitating collisionless system. Appendix~\ref{sec:statistics2} provides a complementary derivation based on Hamilton's equation,
extending \cite{Binney1988} to  self gravitating systems. Section~\ref{sec:2dcaseWKB} focuses on razor thin axisymmetric galactic discs within the WKB approximation.
 Some of the underlying calculations are postponed to  Appendixes~\ref{sec:responsematrixcoefficients} and~\ref{sec:autocorrelationdiagonalization}. Finally,  section~\ref{sec:conclusion} wraps up.

\section{Secular diffusion equation}
\label{sec:seculardiffusionequation}

The secular diffusion equation aims at describing the long-term aperiodic evolution of a self-gravitating collisionless system, perturbed by exterior potential fluctuations. A typical application for this formalism is the study of a galactic disc undergoing (cosmic) perturbations from  its surrounding dark matter halo or the  secular diffusion of accretion streams within the Galactic halo. We will suppose that the background gravitational potential of the system is stationary and integrable, so that we may always remap the usual phase-space coordinates $(\bm{x}, \bm{v})$ to the angle-action coordinates $(\bm{\theta} , \bm{J})$. This is a strong assumption,
as one could imagine situations where the secular evolution breaks symmetry warranting integrability. The angles $\bm{\theta}$ are $2 \pi-$periodic,
 whereas the actions $\bm{J}$ are conserved for a few dynamical times and secularly drift with cosmic time.

\subsection{Evolution equations}
\label{sec:evolutionequations}

We consider a stationary Hamiltonian $H_0 (\bm{J})$, associated to a stationary background gravitational potential $\psi_0$. We also consider a quasi-stationary distribution function $F_0 (\bm{J},t)$, which, at fixed secular time, only depends on the actions thanks to Jeans theorem \citep{Jeans1915}. Finally, we suppose that an exterior source is perturbing this stationary system, so that we can expand the distribution function and the Hamiltonian of the system as
\begin{equation}
\begin{cases}
\displaystyle F (\bm{J}, \bm{\theta},t) = F_0(\bm{J},t) + f(\bm{J}, \bm{\theta},t) \, ,
\\  
\displaystyle H(\bm{J}, \bm{\theta}, t) = H_0(\bm{J}) + \psi^e(\bm{J},\bm{\theta},t) + \psi^s(\bm{J},\bm{\theta},t) \, ,
\end{cases}
\label{perturbation}
\end{equation}
where $f$ is the perturbation of the distribution function, $\psi^e$ is the perturbing \textit{exterior} potential generated by the exterior source, and $\psi^s$ is the \textit{self-response} from the system induced by its self-gravity. This decomposition now involves  two main temporal scales. The shortest scale is the fluctuation timescale, during which $F_0(\bm{J})$ may be considered constant. The longest timescale corresponds to the secular evolution timescale. The perturbations are supposed to be small so that ${f \ll F_{0}}$ and ${\psi^{e} , \psi^{s} \ll \psi_{0}}$. The evolution of the collisionless system is then driven by Boltzmann collisionless equation which reads
\begin{equation}
\frac{\mathrm{d} F}{\mathrm{d} t} = \frac{\partial F}{\partial t} + \left\{ H , F \right\} = 0 \, ,
\label{boltzmann_equation}
\end{equation}
where $\{H , F\} $ is the Poisson bracket. Injecting the decomposition from equation~\eqref{perturbation} in Boltzmann's equation~\eqref{boltzmann_equation}, we obtain
\begin{align}
 \displaystyle 0 = & \frac{\partial F_0}{\partial t}+ \frac{\partial f}{\partial t}  - \!\left[ \frac{\partial \psi^e}{\partial \bm{\theta}}  \!+\! \frac{\partial \psi^s}{\partial \bm{\theta}}\right]  \!\cdot\! \frac{\partial F_0}{\partial \bm{J}} - \!\left[ \frac{\partial \psi^e}{\partial \bm{\theta}} \!+\! \frac{\partial \psi^s}{\partial \bm{\theta}}\right] \!\cdot\! \frac{\partial f}{\partial \bm{J}} \nonumber
\\
\label{injection_equa_diff} & \;\; +\displaystyle  \bm{\Omega} \!\cdot\! \frac{\partial f}{\partial \bm{\theta}} + \left[ \frac{\partial \psi^e}{\partial \bm{J}} \!+\! \frac{\partial \psi^s }{\partial \bm{J}}\right] \!\cdot\! \frac{\partial f}{\partial \bm{\theta}}\, ,
\end{align}
where we have defined the frequencies of the motion on the action-torii as 
\begin{equation}
\dot{\bm{\theta}} = \bm{\Omega} = \frac{\partial {H_0}}{\partial \bm{J}} \, .
\label{def_omega}
\end{equation}
In order to derive the corresponding secular equation, we perform an angle-average on $\bm{\theta}$ of equation~\eqref{injection_equa_diff}. All terms involving a single derivation $\partial / \partial \bm{\theta}$ are equal to $0$, since the angles $\bm{\theta}$ are $2 \pi$-periodic. Moreover, we have ${\int_{\bm{\theta}} \mathrm{d} \bm{\theta} \, f = 0}$, because all the variations independant of $\bm{\theta}$ are included in $F_{0}(\bm{J},t)$. As $F_0$ is independent of $\bm{\theta}$, we obtain
\begin{align}
\frac{\partial F_0}{\partial t} = & \frac{1}{(2 \pi)^{d}} \int \! \mathrm{d} \bm{\theta} \, \left[ \frac{\partial \psi^e}{\partial \bm{\theta}} \!+\! \frac{\partial \psi^s}{\partial \bm{\theta}}\right] \!\cdot\! \frac{\partial f}{\partial \bm{J}} \, \nonumber
\\
\label{secular_boltzmann_calculation}  - & \frac{1}{(2 \pi)^{d}} \int \! \mathrm{d} \bm{\theta} \, \left[ \frac{\partial \psi^e}{\partial \bm{J}} \!+\! \frac{\partial \psi^s }{\partial \bm{J}}\right] \!\cdot\! \frac{\partial f}{\partial \bm{\theta}} \, ,
\end{align}
where $d$ is the dimension of the physical space. Using Schwartz theorem, this secular diffusion equation can be written under the shorter form
\begin{equation}
\frac{\partial F_0}{\partial t} = \frac{1}{(2 \pi)^{d}} \frac{\partial }{\partial \bm{J}} \!\cdot\! \left[ \int \! \mathrm{d} \bm{\theta} \, f \, \frac{\partial [ \psi^{e} \!+\! \psi^{s} ] }{\partial \bm{\theta}}\right] \, .
\label{secular_boltzmann}
\end{equation}
Equation~\eqref{secular_boltzmann}, written as the divergence of a flux, emphasizes the fact that during  orbital diffusion the total number of stars is strictly conserved. Recalling that ${f \ll F_0}$ and ${\psi^e,\psi^s \ll \psi_0}$, the secular evolution equation~\eqref{secular_boltzmann} shows that $\partial F_0 / \partial t$ is in fact a second order term. 

Correspondingly, keeping only first order-terms in~\eqref{injection_equa_diff}, we obtain the second diffusion equation  for the short timescale, which reads
\begin{equation}
\frac{\partial f}{\partial t} + \bm{\Omega} \!\cdot\! \frac{\partial f}{\partial \bm{\theta}} - \frac{\partial F_0}{\partial \bm{J}} \!\cdot\! \frac{\partial [ \psi^e \!+\! \psi^s ]}{\partial \bm{\theta}} = 0 \, .
\label{linearization}
\end{equation}
This equation describes the evolution of the perturbation distribution function $f$ on the fast fluctuating timescale. On such timescales, $ \partial F_0 / \partial \bm{J} $ will be considered as independant of $t$. The next step is to study the fast fluctuating equation~\eqref{linearization}, whose solutions will allow us to estimate the diffusion coefficients for the secular evolution given by equation~\eqref{secular_boltzmann} and describe the diffusion of the quasi-stationary distribution function $F_{0}$ in action-space.

\subsection{Fourier expansion}
\label{sec:fourierexpansion}

One of the many assets of the angle-action variables is that the angles $\bm{\theta}$ are $2 \pi-$periodic allowing us to perform discrete Fourier expansions with respect to these variables. We define the Fourier transform in angles of a function $X (\bm{\theta} , \bm{J})$ as
\begin{equation}
\begin{cases}
\displaystyle X (\bm{\theta} , \bm{J}) = \! \sum\limits_{\bm{m} \in \mathbb{Z}^{d}} \!\! X_{\bm{m}} (\bm{J}) \, e^{i \bm{m} \cdot \bm{\theta}} \, ,
\\
\displaystyle X_{\bm{m}} (\bm{J}) = \frac{1}{(2 \pi)^{d}} \! \int \!\! \mathrm{d} \bm{\theta} \, X (\bm{\theta} , \bm{J}) \, e^{- i \bm{m} \cdot \bm{\theta}} \, .
\end{cases} 
\label{definition_Fourier}
\end{equation}
Thanks to this transformation, the evolution equation~\eqref{linearization} takes the form
\begin{equation}
\frac{\partial f_{\bm{m}}}{\partial t} + i \bm{m} \!\cdot\! \bm{\Omega} \, f_{\bm{m}} - i \bm{m} \!\cdot\! \frac{\partial F_0}{\partial \bm{J}} \,\left[ \psi_{\bm{m}}^e \!+\! \psi_{\bm{m}}^s \right] = 0 \, .
\label{linearization_boltzmann}
\end{equation}
At this stage, we introduce the assumption of timescale decoupling and \textit{push} the secular time to infinity. As a consequence, in the upcoming calculations, we will suppose that ${\partial F_{0} / \partial \bm{J} = cst.}$ Forgetting transient terms and bringing the initial time to $- \infty$, to focus only on the forced regime, the equation~\eqref{linearization_boltzmann} can be solved explicitly, leading to
\begin{equation}
f_{\bm{m}} (\bm{J} , t) \! = \!\!\int_{- \infty}^{t} \!\!\!\!\!\! \mathrm{d} \tau e^{- i \bm{m} \cdot \bm{\Omega} (t - \tau)} i \bm{m} \!\cdot\! \frac{\partial F_0}{\partial \bm{J}} \left[ \psi^e_{\bm{m} } \!+\! \psi^s_{\bm{m}}  \right] \!(\bm{J} , \tau) \, .
\label{solution_linearization}
\end{equation}
We define the temporal Fourier transform as
\begin{equation}
\begin{cases}
\displaystyle \widehat{f} (\omega) = \int_{- \infty}^{+ \infty} \!\!\!\!\!\!\mathrm{d} t \, f (t) \, e^{ i \omega t} \, ,
\\
\displaystyle f (t) =  \frac{1}{2 \pi} \int_{- \infty}^{+ \infty} \!\!\!\!\!\! \mathrm{d} \omega \, \widehat{f} (\omega) \, e^{- i \omega t} \, .
\end{cases}
\label{definition_Fourier_temporal}
\end{equation}
Taking the Fourier transform of equation~\eqref{linearization_boltzmann} at the frequency $\omega$, one can write
\begin{equation}
\widehat{f}_{\bm{m}} (\bm{J} , \omega) = - \frac{ \bm{m} \!\cdot\!  \partial F_{0} / \partial \bm{J} }{\omega \!-\! \bm{m} \!\cdot\! \bm{\Omega}} \left[ \widehat{\psi^{e}}_{\!\!\!\bm{m}} (\bm{J} , \omega) + \widehat{\psi^{s}}_{\!\!\!\bm{m}} (\bm{J} , \omega) \right] \, .
\label{linearization_boltzmann_Fourier}
\end{equation}

\subsection{Matrix method}

An important property of this self gravitating system is that the perturbed distribution function $f$ is consistent with the self-gravitating potential $\psi^{s}$ and its associated density $\rho^{s}$, so that we have
\begin{equation}
\rho^{s} (\bm{x} ,t) = \int \mathrm{d} \bm{v} \, f (\bm{x}, \bm{v}, t) =  \sum_{\bm{m}} \int \! \mathrm{d} \bm{v} \, f_{\bm{m}} (\bm{J},t) \, e^{i \bm{m} \cdot \bm{\theta}} \, .
\label{link_rho_s_small_f}
\end{equation}
In order to simplify further equation~\eqref{link_rho_s_small_f}, we follow Kalnajs matrix method \citep{Kalnajs2} and introduce a complete biorthonormal basis of potentials and densities ${\psi^{(p)} (\bm{x})}$ and ${\rho^{(p)} (\bm{x})}$, such that
\begin{equation}
\begin{cases}
\displaystyle \nabla ^{2} \psi^{(p)} = 4 \pi G \rho^{(p)} \, ,
\\
\displaystyle \int \! \mathrm{d} \bm{x} \, [{\psi^{(p)}} (\bm{x})]^{*} \, \rho^{(q)} (\bm{x}) = - \delta_{p}^{q} \, .
\end{cases}
\label{definition_basis}
\end{equation}
On such a basis, we can write the exterior and the self potentials as
\begin{equation}
\begin{cases}
\displaystyle \psi^{s} (\bm{x}, t) = \sum_{p \in \mathbb{N}}{a_{p} (t) \, \psi^{(p)} (\bm{x})} \, ,
\\
\displaystyle \psi^{e} (\bm{x}, t) = \sum_{p \in \mathbb{N}}{b_{p} (t) \, \psi^{(p)}(\bm{x})} \, . 
\end{cases}
\label{basis_potential_development}
\end{equation}
The linearity of Poisson's equation ensures that this decomposition translates into ${\rho^{s} (\bm{x} , t) \!=\! \sum_{p} \! a_{p} (t) \rho^{(p)} (\bm{x})}$. Using the biorthogonality of the basis, we multiply equation~\eqref{link_rho_s_small_f} by ${[ \psi^{(p)} (\bm{x})]^{*}}$ and integrate over all positions to obtain
\begin{equation}
a_p (t) = - \!\sum_{\bm{m}}  \!\int \! \mathrm{d} \bm{x} \, \mathrm{d} \bm{v} \, f_{\bm{m}} (\bm{J} , t) \, e^{i \bm{m} \cdot \bm{\theta}} \, [\psi^{(p)} (\bm{x})]^{*} \, .
\label{estimation_a_II}
\end{equation}

\subsection{Response matrix and self-consistency}

As the transformation ${(\bm{x},\bm{v}) \mapsto (\bm{\theta}, \bm{J})}$ is canonical, we have ${\mathrm{d} \bm{x} \, \mathrm{d} \bm{v} = \mathrm{d} \bm{\theta} \, \mathrm{d} \bm{J}}$. The integration on $\bm{\theta}$ in equation~\eqref{estimation_a_II} is straigthforward since only ${[ \psi^{(p)} (\bm{x})]^{*} e^{i \bm{m} \cdot \bm{\theta}}}$ depends on it, so that it becomes
\begin{equation}
a_p(t) = - (2 \pi)^{d} \sum_{\bm{m}} \!\int \!\! \mathrm{d} \bm{J}  \, f_{\bm{m}} (\bm{J} , t) \, [\psi^{(p)}_{\bm{m}} (\bm{J})]^{*} \, .
\label{estimation_a_III}
\end{equation}
Using the expression~\eqref{linearization_boltzmann_Fourier} and taking the temporal Fourier transform of equation~\eqref{estimation_a_III} at the frequency $\omega$, one obtains
\begin{align}
\widehat{a}_{p} (\omega) = & \, (2 \pi)^{d} \sum_{q} \left[ \widehat{a}_{q} (\omega) + \widehat{b}_{q} (\omega) \right] \, \times \nonumber
\\
\label{estimation_a_IV} & \sum_{\bm{m}} \!\int \!\! \mathrm{d} \bm{J} \, \frac{ \bm{m} \!\cdot\! \partial F_{0} / \partial \bm{J} }{\omega \!-\! \bm{m} \!\cdot\! \bm{\Omega}} \, [\psi^{(p)}_{\bm{m}} (\bm{J})]^{*} \, \psi^{(q)}_{\bm{m}} (\bm{J}) \, .
\end{align}
We define the response matrix of the system $\widehat{\mathbf{M}}$ as
\begin{equation}
\!\!\!\! \widehat{\mathbf{M}}_{p q} (\omega) \!=\! (2 \pi)^{d} \!\sum\limits_{\bm{m}} \!\int \! \mathrm{d} \bm{J}  \frac{\bm{m} \!\cdot\! \partial F_{0} \!/\! \partial \bm{J} }{\omega \!-\! \bm{m} \!\cdot\! {\Omega}} [ \psi^{(p)}_{\bm{m}} (\bm{J}) ]^{*} \, \psi^{(q)}_{\bm{m}} (\bm{J}) \, , \!\!
\label{Fourier_M}
\end{equation}
where one must note that the response matrix depends only on the initial equilibrium state of the disc, since ${\partial F_{0} / \partial \bm{J}}$ evolves only on the secular scale, the perturbing and self-gravitating potentials are absent, and the basis elements $\psi^{(p)}$ are chosen once for all. Finally, in order to shorten the notations, the amplitudes of the self and exterior potentials are defined as ${\bm{a} (t) \!=\! (a_{1} (t) , ... , a_{p} (t) , ...)}$ and ${\bm{b} (t) \!=\! (b_{1} (t) , ... , b_{p} (t) , ... )}$. Thanks to these notations, one can simplify equation~\eqref{estimation_a_IV}, and rewrite it under the form
\begin{equation}
\widehat{\bm{a}} (\omega) \!+\! \widehat{\bm{b}} (\omega) = \left[ \mathbf{I} \!-\! \widehat{\mathbf{M}} (\omega) \right]^{-1} \!\!\!\cdot \widehat{\bm{b}} (\omega) \, .
\label{estimation_a_Laplace}
\end{equation}
One should note that the matrix ${ [\mathbf{I} \!-\! \widehat{\mathbf{M}}] }$ is invertible only if the self-gravitating system is linearly stable so that all the  eigenvalues of $\widehat{\mathbf{M}}$ are assumed to be strictly smaller than 1 for all values of $\omega$.

\subsection{Diffusion coefficients}
\label{sec:diffusioncoefficients}

The amplification relation~\eqref{estimation_a_Laplace} corresponds to the short timescale (dynamical) response of the system, driven by the evolution equation~\eqref{linearization}. We will now describe the impact of these solutions on the long timescale diffusion equation given by equation~\eqref{secular_boltzmann}.  Starting from equation~\eqref{secular_boltzmann},
one has to evaluate an expression of the form
\begin{align} 
\label{new_way_diffusion_coefficients} & \!\!\!\!\!\! \displaystyle\frac{1}{(2 \pi)^{d}} \!\int \!\! \mathrm{d} \bm{\theta} \, f (\bm{J},\bm{\theta},t) \, \frac{\partial \left[ \psi^{e} \!+\! \psi^{s} \right]}{\partial \bm{\theta}} = 
\\
& \;\;\;\displaystyle \frac{1}{(2 \pi)^{d}} \!\!\sum\limits_{\bm{m}_{1} , \bm{m}_{2}} \!\!\int \!\! \mathrm{d} \bm{\theta} \,  f_{\bm{m}_1} i \bm{m}_{2} \left[ \psi^{e}_{\bm{m}_2} \!+\! \psi^{s}_{\bm{m}_2} \right] e^{i (\bm{m}_{1} + \bm{m}_{2}) \cdot \bm{\theta}} \, . \nonumber
\end{align}
Here, only  terms with ${ \bm{m}_{1} = - \bm{m}_{2} }$ are different from $0$. Using equation~\eqref{solution_linearization} and the fact that ${ \psi_{- \bm{m}} = \psi^{*}_{\bm{m}} }$, we can finally rewrite the diffusion equation~\eqref{secular_boltzmann} under the form
\begin{equation}
\frac{\partial F_0}{\partial t} = \sum_{\bm{m}} \bm{m} \!\cdot\! \frac{\partial}{\partial \bm{J}} \left[ D_{\bm{m}} (\bm{J}) \, \bm{m} \!\cdot\! \frac{\partial F_0}{\partial \bm{J}} \right] \, ,
\label{diffusion_equation_II}
\end{equation}
where the anisotropic diffusion coefficients $D_{\bm{m}} (\bm{J})$ are given by
\begin{equation}
\begin{aligned}
\hskip -0.5cm
D_{\bm{m}} (\bm{J}, t) = & \left[ \psi^{e \, *}_{\bm{m}} (\bm{J},t) \!+\! \psi^{s \, *}_{\bm{m}} (\bm{J},t) \right] \, \times
\\
& \int_{- \infty}^{t} \!\!\!\!\!\! \mathrm{d} \tau \, e^{- i \bm{m} \cdot \bm{\Omega} (t - \tau)} \left[  \psi^{e}_{\bm{m} } (\bm{J},\tau) \!+\! \psi^{s}_{\bm{m}} (\bm{J}, \tau) \right] \, .
\end{aligned}
\label{definition_D}
\end{equation}
Note that equation~\eqref{diffusion_equation_II} can be re-arranged as 
\begin{equation}
\frac{\partial F_0}{\partial t} \!=\! \frac{\partial}{\partial \bm{J}} \!\cdot\! \left[ \mathbf{ D} (\bm{J}) \!\cdot\!
\frac{\partial F_0}{\partial \bm{J}} \right] ,\,
{\rm with}\,\,
\mathbf{ D}(\bm{J}) \!=\! \sum_{\bm{m}} \!D_{\bm{m}} (\bm{J}) \, \bm{m}  \!\otimes\! \bm{m}\,, \nonumber
\end{equation}
an anisotropic tensor diffusion matrix. Using the basis decomposition introduced in equation~\eqref{basis_potential_development}, the diffusion coefficients from equation~\eqref{definition_D} take the form
\begin{align}
D_{\bm{m}} (\bm{J}, t) = & \sum_{p , q} \psi^{(p)}_{\bm{m}} \psi^{(q) *}_{\bm{m}} \big [a_q^{*} (t) \!+\! b_q^{*} (t) \big]  \, \times \nonumber
\\
\label{expression_D} & \int_{- \infty}^{t} \!\!\!\! \mathrm{d} \tau \, e^{- i \bm{m} \cdot \bm{\Omega} (t - \tau)} \big[ a_p (\tau) \!+\! b_p (\tau) \big] \, .
\end{align}
Expressing the temporal coefficients $a_{p} (t)$ and $b_{p} (t)$ via their Fourier transforms, we obtain
\begin{align}
\!\!\!\! D_{\bm{m}} (\bm{J}, t) = & \frac{1}{(2 \pi)^{2}} \sum_{p , q} \psi^{(p)}_{\bm{m}} \psi^{(q) *}_{\bm{m}} \!\!\!\int \!\! \mathrm{d} \omega \big[a_q^{*}  \!+\! b_q^{*} \big] (\omega)  \, e^{i \omega t} \, \times \nonumber
\\
\label{expression_D_II} & \! \int_{- \infty}^{t} \!\!\!\!\!\! \mathrm{d} \tau \, e^{- i \bm{m} \cdot \bm{\Omega} (t - \tau)} \!\!\!\int \!\! \mathrm{d} \omega '  \big[ a_p \!+\! b_p \big] (\omega') \, e^{- i \omega ' \tau } \, .
\end{align}
The amplification relation~\eqref{estimation_a_Laplace} allows us to rewrite equation~\eqref{expression_D_II} as
\begin{align}
\!\!\!\! D_{\bm{m}} (\bm{J}, t) = & \frac{1}{(2 \pi)^{2}} \sum_{p , q} \sum_{p_{1} , q_{1}} \psi^{(p)}_{\bm{m}} \psi^{(q) *}_{\bm{m}}  \, \times \nonumber
\\
\label{expression_D_III} & \!\!\!\!\!\!\! \int \!\! \mathrm{d} \omega \, e^{i \omega t}  \left[ \big[ \mathbf{I} \!-\! \widehat{\mathbf{M}} (\omega) \big]^{-1}_{q q_{1}} \right]^{*} \, \widehat{b}_{q_{1}}^{*} (\omega) \, \times 
\\
& \!\!\!\!\!\!\! \int_{- \infty}^{t} \!\!\!\!\!\!\! \mathrm{d} \tau \, e^{- i \bm{m} \cdot \bm{\Omega} (t - \tau)} \!\!\!\int \!\! \mathrm{d} \omega ' e^{- i \omega ' \!\tau } \big[ \mathbf{I} \!-\! \widehat{\mathbf{M}} (\omega') \big]^{-1}_{p p_{1}} \, \widehat{b}_{p_{1}} (\omega') \, . \nonumber
\end{align}

\subsection{Statistical expectation}
\label{sec:statisticalapproach}

The final stage of the  derivation is to introduce the statistics  of the external perturbations. Indeed, our previous calculation corresponds to the response 
of the system  to a {\sl given} particular perturbation history: ${ t \mapsto \bm{b} (t) }$. Let us now denote the ensemble average operation on such different realizations as ${\langle\,.\,\rangle }$. As the global underlying background gravitational potential is assumed to be stationary, the mapping ${ (\bm{x} , \bm{v}) \mapsto (\bm{\theta} , \bm{J}) }$ remains the same for the different realizations, so that the operations of derivation or integration with respect to $\bm{\theta}$ and $\bm{J}$ commute with the ensemble average. The diffusion equation~\eqref{diffusion_equation_II}, when ensemble averaged, takes the form
\begin{equation}
\frac{\partial \langle F_{0} \rangle}{\partial t} = \sum\limits_{\bm{m}} \bm{m} \!\cdot\! \frac{\partial }{\partial \bm{J}} \!\left[\left\langle D_{\bm{m}} (\bm{J}) \, \bm{m} \!\cdot\! \frac{\partial F_{0}}{\partial \bm{J}}\right\rangle\right] \, .
\label{averaged_diffusion_equation}
\end{equation}
\textit{A priori}, the gradient $\partial F_{0} / \partial \bm{J}$, cannot be taken out of the ensemble average operation. However, we intend to describe the effect of an averaged fluctuation on a given $F_{0}$ representing a mean disc. Then, one may assume that the quasi-stationary distribution function $F_{0}$, its gradients and therefore the response matrix $\widehat{\mathbf{M}}$ do not change significantly from one realization to another, so that they can be taken out of the ensemble average operation. This means that we assume there exists a mean response for the secular distribution, 
${ F_{0}=\langle F_0 \rangle }$, de-correlated from the perturbations, so that we have ${ \langle D_{\bm{m}} (\bm{J}) \, \bm{m} \!\cdot\! \partial F_{0} / \partial \bm{J} \rangle = \langle D_{\bm{m}} (\bm{J}) \rangle \, \bm{m} \!\cdot\! \partial F_{0} / \partial \bm{J} }$. We also suppose that the time evolution of the exterior perturbing potential  is stationary  and therefore introduce the corresponding temporal autocorrelation function defined as
\begin{equation}
\mathbf{C}_{k l} (t_{1} \!-\! t_{2}) =\langle b_{k} (t_{1}) \, b_{l}^{*} (t_{2})\rangle \, ,
\label{stationarity_hypothesis}
\end{equation}
where the exterior perturbation is also assumed to be of zero mean. The autocorrelation $\mathbf{C}$ connects the temporal coefficients $\bm{b}$, whereas the diffusion coefficients from equation~\eqref{expression_D_III} involve the Fourier transformed ones $\widehat{\bm{b}}$, so that one needs to compute ${ \big< \widehat{b}_{k} (\omega) \, \widehat{b}_{l}^{*} (\omega') \big> }$. One can straightforwardly show that 
\begin{equation}
\big< \widehat{b}_{k} (\omega ) \, \widehat{b}_{l}^{*} (\omega ' ) \big> = 2 \pi \, \delta_{\rm D} (\omega \!-\! \omega') \, \widehat{\mathbf{C}}_{k l} (\omega) \, ,
\label{autocorrelation_Fourier_II}
\end{equation}
where $\widehat{\mathbf{C}}$ is the temporal Fourier transform of the autocorrelation of the external potential.
Using this result in the ensemble averaged expression~\eqref{expression_D_III} yields
\begin{align}
\langle D_{\bm{m}} (\bm{J} , t)\rangle =  \frac{1}{2 \pi} \sum_{p , q} & \psi^{(p)}_{\bm{m}} \psi^{(q) *}_{\bm{m}} \!\!\int \! \mathrm{d} \omega \!\int_{- \infty}^{0} \!\!\!\!\!\! \mathrm{d} \tau' \, e^{- i (\omega - \bm{m} \cdot \bm{\Omega}) \tau '} \nonumber
\\
& \!\!\!\! \label{expression_D_IV} \!\!\left[ [ \mathbf{I} \!-\! \widehat{\mathbf{M}} ]^{-1} \!\!\cdot \widehat{\mathbf{C}} \cdot [ \mathbf{I} \!-\! \widehat{\mathbf{M}} ]^{-1}\right]_{pq} \!(\omega) \, ,
\end{align}
where we performed the change of variables ${ \tau ' = \tau \!-\! t }$. One should note that when ensemble averaged, the diffusion coefficients are (explicitly) independent of $t$\footnote{though they depend on the secular timescale via the variation of $F$ in $\mathbf{\widehat M}$.}. In order to shorten temporarily the notations, we introduce ${ \widehat{\mathbf{L}} = [\mathbf{I} \!-\! \widehat{\mathbf{M}}]^{-1} \!\cdot\! \widehat{\mathbf{C}} \!\cdot\! [\mathbf{I} \!-\! \widehat{\mathbf{M}}]^{-1} }$. In equation~\eqref{expression_D_IV}, one has to evaluate an expression of the form
\begin{equation}
\frac{1}{2 \pi} \!\int_{- \infty}^{+ \infty} \!\!\!\!\!\!\!\!\! \mathrm{d} \omega \, \widehat{\mathbf{L}} (\omega) \!\!\int_{- \infty}^{0} \!\!\!\!\!\!\!\mathrm{d} \tau' e^{- i (\omega - \bm{m} \cdot \bm{\Omega}) \tau'} \!\!=\! \frac{i}{2 \pi} \!\!\int_{- \infty}^{+ \infty} \!\!\!\!\!\!\!\!\! \mathrm{d} \omega \, \frac{\widehat{\mathbf{L}}(\omega)}{\omega \!-\! \bm{m} \!\cdot\! \bm{\Omega}} \, ,
\label{shape_integration_pv}
\end{equation}
where in the integration over $\tau'$ we only kept the term for ${ \tau' = 0 }$, by adding an imaginary part to the frequency $\omega$ so that ${ \omega = \omega \!+\! i 0^{+} }$, ensuring the convergence for ${ \tau' \!\to\! - \infty }$. The remaining integral over $\omega$ can be evaluated using Plemelj formula
\begin{equation}
\frac{1}{x \pm i 0^{+}} = \mathcal{P} \!\left( \frac{1}{x} \right) \mp i \pi \delta_{\rm D} (x) \, ,
\label{Plemelj_formula}
\end{equation}
where $\mathcal{P}$ denotes Cauchy principal value. Therefore, equation~\eqref{shape_integration_pv} becomes
\begin{align}
\eqref{shape_integration_pv} \propto
 \frac{i}{2 \pi} \mathcal{P} \!\! \int_{- \infty}^{+ \infty} \!\!\!\!\!\!\!\! \mathrm{d} \omega \, \frac{\widehat{\mathbf{L}} (\omega)}{\omega \!-\! \bm{m} \!\cdot\! \bm{\Omega}} +
\label{shape_integration_pv_II} 
\frac{1}{2} \, \widehat{\mathbf{L}} (\bm{m} \!\cdot\! \bm{\Omega}) \, .
\end{align}
The final step of the derivation is to show that the principal value term present in equation~\eqref{shape_integration_pv_II} has no impact on the secular diffusion. Indeed, using the expression~\eqref{definition_D} of the diffusion coefficients, one can show that they satisfy ${ D_{- \bm{m}} (\bm{J}) = D_{\bm{m}}^{*} (\bm{J}) }$. As a consequence, as we are summing on all the modes $\bm{m}$, the diffusion equation~\eqref{diffusion_equation_II} may be rewritten under the form
\begin{equation}
\frac{\partial F_{0}}{\partial t} = \sum\limits_{\bm{m}} \bm{m} \!\cdot\! \frac{\partial }{\partial \bm{J}} \!\left[ \text{Re} \left[ D_{\bm{m}} (\bm{J}) \right] \, \bm{m} \!\cdot\! \frac{\partial F_{0}}{\partial \bm{J}} \right] \, .
\label{diffusion_equation_II_real}
\end{equation}
From equations~\eqref{Fourier_M} and~\eqref{stationarity_hypothesis}, we know that the response matrix and the autocorrelation matrix are hermitian so that ${ \widehat{\mathbf{M}}^{*} \!= \widehat{\mathbf{M}}^{t} }$ and ${ \widehat{\mathbf{C}}^{*} \!= \widehat{\mathbf{C}}^{t} }$.  As a consequence, the matrix $\widehat{\mathbf{L}}$ is also hermitian. Since ${\text{Re} (D_{\bm{m}}) \!=\! (D_{\bm{m}} \!+\! D_{\bm{m}}^{*})/ 2}$, starting from equation~\eqref{shape_integration_pv_II}, we immediately obtain
\begin{align}
\langle \text{Re} \!\left[ D_{\bm{m}} (\bm{J}) \right]\rangle = & \, \frac{1}{2} \sum_{p , q} \!\psi^{(p)}_{\bm{m}} \psi^{(q)*}_{\bm{m}} \times \nonumber
\\
& \label{expression_Re_Dm_averaged} \left[ [\mathbf{I} \!-\! \widehat{\mathbf{M}}]^{-1} \!\!\cdot \widehat{\mathbf{C}} \cdot\! [ \mathbf{I} \!-\! \widehat{\mathbf{M}} ]^{-1} \!\right]_{pq} \!\!\!(\bm{m} \!\cdot\! \bm{\Omega}) \, ,
\end{align}
so that the full secular diffusion equation takes the form
\begin{align}
\frac{\partial F_0}{\partial t} = \sum_{\bm{m}} & \, \bm{m} \!\cdot\! \frac{\partial}{\partial \bm{J}} \bigg[  \bm{m} \!\cdot\! \frac{\partial F_0}{\partial \bm{J}}  \sum_{p , q} \frac{1}{2} \, \psi^{(p)}_{\bm{m}} (\bm{J}) \, \psi^{(q) *}_{\bm{m}} (\bm{J}) \, \times \nonumber
\\
\label{diffusion_equation_statistical} & \;\;\;\;\;\;\left[[ \mathbf{I} \!-\! \widehat{\mathbf{M}}]^{-1} \!\cdot \, \widehat{\mathbf{C}} \cdot\! \,  [\mathbf{I} \!-\! \widehat{\mathbf{M}}]^{-1} \!\right]_{p q} \!\!\!( \bm{m} \!\cdot\! \bm{\Omega})  \bigg] \, .
\end{align}
Equation~(\ref{diffusion_equation_statistical}) is the main result of this section.
In Appendix~\ref{sec:statistics2}, we present an alternative derivation of these diffusion coefficients based on Hamilton's equations from which we recover the exact same diffusion equation. The ${ 1/2 }$ factor recovered via these two complementary approaches was skipped in the calculation presented in \cite{Pichon2006}, because of an error in the bounds of half-temporal integrations, similar to the one present in equation~\eqref{shape_integration_pv}.
This derivation is valid in any dimensions, provided the underlying system is integrable.
One may also note that in the homogeneous limit, equation~\eqref{diffusion_equation_statistical} reduces to the secular diffusion equation obtained in~\cite{Chavanis2012EPJ,Chavanis2013b}.
In the next section we will restrict ourselves to 2D configurations 
and make further assumptions in order to simplify equation~\eqref{expression_Re_Dm_averaged} into a one dimensional quadrature.

\section{Thin tepid discs and WKB limit}
\label{sec:2dcaseWKB}

One difficulty for the implementation of the secular diffusion equation~\eqref{diffusion_equation_statistical} is to simultaneously have an explicit mapping ${(\bm{x},\bm{v}) \!\mapsto\! (\bm{\theta},\bm{J})}$ to the angle-actions coordinates, and be able to evaluate the diffusion coefficients given by equation~\eqref{expression_Re_Dm_averaged}, which require to invert the response matrix ${ [\mathbf{I} \!-\! \widehat{\mathbf{M}}] }$. In order to deal with the \textit{non-locality} of Poisson's equation, we also have to explicitly introduce potential basis elements, $\psi^{(p)}$, as in equation~\eqref{definition_basis}, to compute the response matrix from equation~\eqref{Fourier_M}. To ease these calculations in a $2D$ axisymmetric disc, one may rely on the WKB assumption \citep{WKB,Toomre1964,Kalnajs1965,Lin1966,Palmer1989}, which assumes that the perturbations and self-responses will take the form of tightly wound spirals, which in turn allows us to write Poisson's equation \textit{locally}. Considering only such perturbations sums up to introducing basis elements with specific properties as detailed later on.

\subsection{Epicyclic approximation and isothermal DF}
\label{sec:epicyclicapproximation}
In order to explicitly build up a mapping ${(\bm{x} , \bm{v}) \mapsto (\bm{\theta}, \bm{J})}$ for an axisymmetric disc, we assume that the disc is sufficiently cold and therefore rely on the so-called epicyclic approximation.

The natural coordinates for an axisymmetric galactic disc are the polar coordinates $(R, \phi)$, with their associated momenta $(p_{R} , p_{\phi})$. Within such coordinates, the stationary Hamiltonian of the system reads
\begin{equation}
H_{0} (R,\phi,p_R,p_{\phi}) = \frac{1}{2} \!\left[ p_R ^2 \!+\! \frac{p_{\phi} ^2}{R^2} \right] \!+\! \, \psi_0 (R) \, ,
\label{hamiltonian_polar}
\end{equation}
where $\psi_{0}$ is the axisymmetric stationary background potential within the disc. The first action of the system is the angular momentum $J_{\phi}$ defined as
\begin{equation}
J_{\phi} = L_z \equiv\frac{1}{2\pi}\oint \mathrm{d}\phi \,  p_\phi = p_{\phi} = R^2 \dot{\phi} \, .
\label{def_Lz}
\end{equation}
As soon as the value of $J_{\phi}$ is imposed, one obtains a new equation of motion for the $R$ variable given by
\begin{equation}
\ddot{R} = - \frac{\partial \psi_{\rm eff}}{\partial R} \, ,
\label{new_epicycle_R_equation}
\end{equation}
where the effective potential is defined as
\begin{equation}
\psi_{\rm eff} (R) = \psi_0 (R) \!+\! \frac{J_{\phi}^{2}}{2 R^{2}} \, .
\label{def_effective_potential}
\end{equation}
The main idea behind the epicyclic approximation is to approximate the radial motion as an harmonic oscillation. For a given value of $J_{\phi}$, we define implicitly the guiding radius $R_{g}$ as
\begin{equation}
0 = \frac{\partial \psi_{\rm eff}}{\partial R} \bigg|_{R_g} \!\!\!\! = \, \frac{\partial \psi_0}{\partial R} \bigg|_{R_g} \!\!\!\! - \frac{J_{\phi} ^2 }{R_g ^3} \, ,
\label{definition_R_g}
\end{equation}
so that $R_{g} (J_{\phi})$ corresponds to the radius for which stars with an angular momentum equal to $J_{\phi}$ evolve on circular orbits. For a stationary potential, the mapping between $R_{g}$ and $J_{\phi}$ is bijective and unambiguous (up to the sign of $J_{\phi}$). We define the azimuthal frequency ${ \Omega (R_{g}) }$ and the epicyclic frequency ${ \kappa (R_{g}) }$ as
\begin{equation}
\begin{cases}
\displaystyle \Omega^{2} (R_{g}) = \frac{1}{R_{g}} \frac{\partial \psi_{0}}{\partial R} \bigg|_{R_{g}} \!\!\!\!\!= \frac{J_{\phi}^{2}}{R_{g}^{4}} \, ,
\\
\displaystyle \kappa^{2} (R_{g}) = \frac{\partial^{2} \psi_{\rm eff} }{\partial R^{2}} \bigg|_{R_{g}}  \!\!\!\!\!=  \frac{\partial^{2} \psi_{0}}{\partial R^{2}} \bigg|_{R_{g}} \!\!\!\!\!+ 3 \frac{J_{\phi}^{2}}{R_{g}^{4}} \, .
\end{cases}
\label{definition_intrinsic_frequencies}
\end{equation}
A Taylor expansion  at first order  near $R_g$ of equation~\eqref{new_epicycle_R_equation} shows that $R$ satisfies the differential equation ${ \ddot R = - \kappa^{2} (R \!-\! R_{g}) }$, which is the evolution equation of an harmonic oscillator centered on $R_{g}$. We introduce the amplitude $A$ of the radial oscillations and define the radial action $J_{r}$ as
\begin{equation}
J_{r} \equiv\frac{1}{2\pi}\oint \mathrm{d} R \, p_R  = \frac{1}{2} \kappa A^{2} \, .
\label{definition_Jr}
\end{equation}
The case $J_{r} = 0$ corresponds to circular orbits. The larger  $J_{r}$, the wider  the radial oscillations of the star. One should note that within the epicyclic approximation, the two intrinsic frequencies $\Omega$ and $\kappa$ are only function of the angular momentum $J_{\phi}$ and do not depend on the radial action $J_{r}$. Finally, one can show \citep{LyndenBell1972,Palmer1994,BinneyTremaine2008} that the mapping between $(R,\phi,p_{R},p_{\phi})$ and $(\theta_{R},\theta_{\phi}, J_{r} , J_{\phi})$ takes at first order the form
\begin{equation}
\begin{cases}
\displaystyle R = R_g \!+\! A \cos (\theta_{R}) \, ,
\\
\displaystyle \phi  = \theta_{\phi} \!-\! \frac{2 \Omega}{\kappa} \frac{A}{R_g} \sin(\theta_{R}) \, .
\end{cases}
\label{mapping_r_phi_palmer}
\end{equation}
Within this approximation, one can easily parametrize plausible stationary distribution functions for a galactic disc, defined as functions of the actions $(J_{\phi} , J_{r})$. Indeed, we suppose that the stationary distribution $F_{0}$ of the disc is a Schwarzschild distribution function (or locally \textit{isothermal}) given by
\begin{equation}
F_{0} (R_g, J_r) = \frac{ \Omega(R_{g}) \, \Sigma (R_g)}{\pi \, \kappa (R_{g}) \, \sigma_r^2 (R_{g}) } \, \exp \!\left[- \frac{\kappa(R_{g}) \, J_r}{\sigma_r^2(R_g)} \right] \, ,
\label{complete_DF_Schwarzschild}
\end{equation}
where $\Sigma (R_{g})$ is the surface density of the disc and $\sigma_{r}^{2} (R_{g})$, which depends on the position in the disc, encodes the typical radial velocity dispersion of the stars at a given radius. The larger $\sigma_{r}^{2}$, the hotter  the disc and the more stable it is.

\subsection{The WKB basis}
\label{sec:WKBbasis}

In order to use a WKB approach in the secular diffusion equation~\eqref{diffusion_equation_II}, one needs to introduce explicitly a basis of density-potentials from which the WKB hypothesis will follow.

\subsubsection{Definition of the basis elements}
\label{sec:definitionWKBbasis}

\begin{figure}
\begin{center}
\begin{tikzpicture}
\draw [->] [thick] (-0.5,0) -- (7.5,0) ; \draw (7.5,0) node[font = \small, right]{$R$} ;
\draw [->] [thick] (-0.3,-2) -- (-0.3,2) ; \draw (-0.3,2) node[font = \small, above]{$\psi$} ;
\draw [dashed , smooth] plot [domain=0:4] ( \x,{exp(- (\x-2)*(\x-2)*2)} ) ; \draw [dashed , smooth] plot [domain=0:4] ( \x,{-exp(- (\x-2)*(\x-2)*2)} ) ;
\draw [thick, smooth, samples = 100] plot [domain=0:4] ( \x,{exp(- (\x-2)*(\x-2)*2) * cos(1000*(\x-2))} );
\draw [dashed] (2,0) -- (2,-1.3) ; \draw [thin] (2,-0.05) -- (2,0.05) ; \draw (2.05,-1.32) node[font = \small, below]{$R_{0}^{p}$} ; 
\draw [|->] [thin] (2,-1.3) -- (3,-1.3) ; \draw (2.5,-1.35) node[font = \small, above]{$\sigma$} ;
\draw [|->] [thin] (2,1.1) -- (2.35,1.1) ; \draw (2.175,1.1) node[font = \small , above]{$ 1 \!/\! k_{r}^{p}$} ;
\draw [dashed , smooth] plot [domain=3.5:7.5] ( \x,{exp(- (\x-5.5)*(\x-5.5)*2)} ) ; \draw [dashed , smooth] plot [domain=3.5:7.5] ( \x,{-exp(- (\x-5.5)*(\x-5.5)*2)} ) ;
\draw [thick, smooth, samples = 200] plot [domain=3.5:7.5] ( \x,{exp(- (\x-5.5)*(\x-5.5)*2) * cos(2000*(\x-5.5))} );
\draw [dashed] (5.5,0) -- (5.5,-1.3) ; \draw [thin] (5.5,-0.05) -- (5.5,0.05) ; \draw (5.55,-1.32) node[font = \small, below]{$R_{0}^{q}$} ; 
\draw [|->] [thin] (5.5,-1.3) -- (6.5,-1.3) ; \draw (6.0,-1.35) node[font = \small, above]{$\sigma$} ;
\draw [|->] [thin] (5.5,1.1) -- (5.68,1.1) ; \draw (5.675,1.1) node[font = \small , above]{$ 1 \!/\! k_{r}^{q}$} ;
\end{tikzpicture}
\caption{\small{Two WKB basis elements. Each Gaussian is centered around a radius $R_{0}$. The typical extension of the Gaussian is given by the decoupling scale $\sigma$, and they are modulated at the radial frequency $k_{r}$.}}
\label{plot_WKB_basis}
\end{center}
\end{figure}
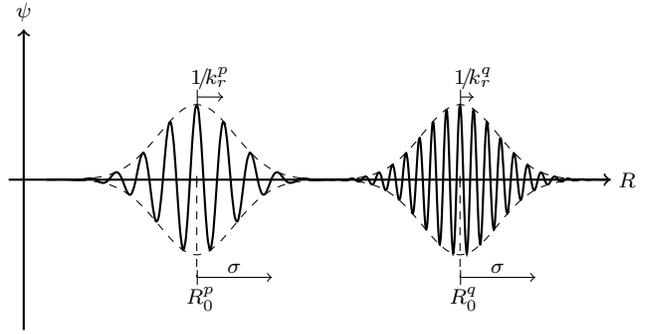

We define on the plane the following basis of potential functions, well suited to represent tightly wound spirals
\begin{equation}
\psi^{[k_{\phi}, k_{r},R_{0}]} (R,\phi) = \mathcal{A} \, e^{ i (k_{\phi} \phi + k_{r} R) } \, \mathcal{B}_{R_{0}} (R) \, ,
\label{definition_tightly_spiral}
\end{equation}
where the functions $\mathcal{B}_{R_{0}} (R)$ are of the form
\begin{equation}
\mathcal{B}_{R_{0}} (R) = \frac{1}{(\pi \sigma^{2})^{1\!/\!4}} \exp \left[ - \frac{(R \!-\! R_{0})^{2}}{2 \sigma^2} \right] \, .
\label{definition_B_WKB}
\end{equation}
To our knowledge, this is the first time that such tightly wound basis elements have been introduced in the context of discs secular dynamics. The central radius $R_{0}$ is the radius around which the Gaussian $\mathcal{B}_{R_{0}}$ is centered, $k_{\phi}$ is an azimuthal number representing the angular component of the basis elements, and $k_{r}$ corresponds to the radial frequency of the potential element. Here $\sigma$ is a scale-separation parameter ensuring the biorthogonality of the basis elements (see below). The radial dependance of the basis elements is illustrated in figure~\ref{plot_WKB_basis}. The amplitude $\mathcal{A}$ is not yet defined and will be chosen in order to guarantee the correct normalization of the basis. The unusual normalization of $\mathcal{B}_{R_{0}}$ will ensure that $\mathcal{A}$ is independent of $\sigma$ and will also allow us to naturally introduce Dirac deltas ${ \delta_{\rm D} (R \!-\! R_{0}) }$ in some of the next calculations.

\subsubsection{Associated surface density elements}
\label{sec:surfacedensityWKB}

Equation~\eqref{definition_basis} requires to have a biorthogonal potential basis. We now determine the surface density basis elements associated to the potentials from equation~\eqref{definition_tightly_spiral}. For 2D razor thin discs, we are in the presence of a discontinuity leading to a surface density $\Sigma (R, \phi)$. In order to satisfy Poisson's Equation, we extend our potential to the $z$-axis via the following Ansatz
\begin{equation}
\psi^{[k_{\phi}, k_{r} , R_{0}]} (R,\phi , z) = \mathcal{A} \, e^{ i (k_{\phi} \phi + k_{r} R) } \, \mathcal{B}_{R_{0}} (R) \, Z(z) \, .
\label{extension_potential_z}
\end{equation}
Injecting such an expression into Poisson's Equation in vacuum ${\Delta \psi^{[ k_{\phi}, k_{r} , R_{0}]} = 0}$, we obtain after some algebra
\begin{align}
\displaystyle-\frac{Z ''}{Z} \!=\!& - k_{r}^{2} \bigg[ 
 1 \!-\! \frac{i}{k_{r} R} \!+\! 2 i \, \frac{R \!-\! R_{0}}{\sigma^{2}} \frac{1}{k_{r}} \!+\! \frac{R \!-\! R_{0}}{R} \frac{1}{(\sigma k_{r})^{2}} + \nonumber
 \\
\label{laplacian_full_extension}  &\;\;\;\;\;\;\;\;\;\displaystyle \frac{1}{(\sigma k_{r})^{2} } \!+\! \frac{k_{\phi}^{2}}{(k_r R)^{2}} \!-\! \left[ \frac{R \!-\! R_{0}}{\sigma^{2}}  \frac{1}{k_r} \right]^{2} \bigg] \, .
\end{align}
In order to obtain a simple expression for the surface density basis elements, we introduce additional WKB-like assumptions, so that the terms appearing in equation~\eqref{laplacian_full_extension} are all negligible in front of $1$. First of all, we assume that the spirals are tightly wound so that we have 
\begin{equation}
k_{r} R \gg 1 \, .
\label{tight_winding_assumption}
\end{equation}
Moreover, introducing the typical size $R_{\rm sys}$ of the system, we add the supplementary constraint
\begin{equation}
k_{r} \, \sigma \gg \frac{R_{\rm sys}}{\sigma} \, .
\label{assumption_gaussian_WKB_rewriting}
\end{equation}
In this limit, assuming that $k_{\phi}$ remains of the order of the unity, equation~\eqref{laplacian_full_extension} becomes
\begin{equation}
\frac{Z ''}{Z} = k_{r}^{2} \, .
\label{equation_Z_extension}
\end{equation}
Therefore, we conclude that within the WKB approximation, the extended 3D potential can be written as
\begin{equation}
\psi^{[ k_{\phi}, k_{r} , R_{0}]} (R, \phi, z) = \psi^{[ k_{\phi}, k_{r} , R_{0}]} (R, \phi) \, e^{- k_r |z|} \, ,
\label{expression_potential_extension}
\end{equation} 
where we added absolute value on $z$ in order to respect the boundaries conditions of the potential at ${z = \pm \infty}$, where the potential has to tend to $0$. Such a potential introduces a discontinuity of ${\partial \psi / \partial z}$ at the plane ${z = 0}$, consistent with the  given surface density. Gauss theorem for the discontinuities at a plane may be written as
\begin{equation}
\Sigma (R,\phi) = \frac{1}{4 \pi G} \left[\lim\limits_{z \to 0^{+}} \frac{\partial \psi}{\partial z} \,  - \lim\limits_{z \to 0^{-}} \frac{\partial \psi}{\partial z} \right] \, .
\label{gauss_theorem_surface}
\end{equation}
We immediately conclude that the surface density associated to a given potential element $\psi^{[ k_{\phi}, k_{r} , R_{0}]}$ is given by
\begin{equation}
\Sigma^{[ k_{\phi}, k_{r} , R_{0}]} (R, \phi) = - \frac{| k_{r} |}{2  \pi G} \, \psi^{[ k_{\phi}, k_{r} , R_{0}]} (R, \phi) \, .
\label{orthogonality_relation_WKB_complex}
\end{equation}

\subsubsection{Biorthogonality condition}
\label{sec:biorthogonalityWKB}

The next step of the definition of the WKB basis is to ensure that this basis is biorthogonal as in equation~\eqref{definition_basis}. Indeed, it has to satisfy the property
\begin{equation}
\delta^{k^{q}_{\phi}}_{k^{p}_{\phi}} \delta^{k^{q}_r}_{k^{p}_r} \delta^{R^{q}_{0}}_{R^{p}_{0}} \!=\!  - \!\!\!\int \!\! \mathrm{d} R  R \,  \mathrm{d} \phi  \!\left[\! \psi^{[ k^{p}_{\phi}, k^{p}_{r} , R^{p}_{0}]} (R, \phi) \!\right]^{*} \!\!\Sigma^{[ k^{q}_{\phi}, k^{q}_{r} , R^{q}_{0}]} (R, \phi) \, .
\label{condition_orthogonality_WKB} \nonumber
\end{equation}
The r.h.s. of this expression becomes
\begin{align}
\label{calcul_condition_orthogonality_WKB}  & \displaystyle 
\frac{| k^{q}_r |}{2 \pi G}  \mathcal{A}_{p} \, \mathcal{A}_{q}  \, \frac{1}{\sqrt{\pi \sigma^{2}} } \!\int \mathrm{d} \phi \, e^{i (k^{q}_{\phi} - k^{p}_{\phi} ) \phi } \,  \, \, \times
\\
&\displaystyle \int \!\mathrm{d} R \, R \,  e^{i (k^{q}_{r} - k^{p}_{r}) R }   \exp \!\left[ - \frac{(R \!-\! R^{p}_0)^{2}}{2 \sigma^2} \right] \exp \!\left[ - \frac{(R \!-\! R^{q}_0)^{2}}{2 \sigma^{2}} \right]  . \nonumber
\end{align}
The integration on $\phi$ is straightforward and gives a term equal to ${ 2 \pi \delta^{k^{p}_{\phi}}_{ k^{q}_{\phi}} }$. In order to be able to integrate on $R$, we now need to impose new WKB-like assumptions to justify the biorthogonality of the basis. We introduce the spatial Fourier transform $\mathcal{F}$ with respect to $R$, and the difference of the two radial frequencies ${ \Delta k_{r} = k^{q}_r \!-\! k^{p}_r }$. Equation~\eqref{calcul_condition_orthogonality_WKB} requires to integrate an expression of the form
\begin{align}
& \!\!\! \displaystyle \int \mathrm{d} R \, e^{i \Delta k_{r} R} \, \mathcal{B}_{R^{p}_{0}} (R) \, \mathcal{B}_{R^{q}_{0}} (R) \nonumber
\\
\label{shape_integration_WKB} & \!\!\! \displaystyle = \int \mathrm{d} k ' \, \mathcal{F} [\mathcal{B}_{R^{p}_{0}}] (k') \, \mathcal{F} [\mathcal{B}_{R^{q}_{0}}] (\Delta k_{r} \!-\! k')
\\
& \!\!\! \displaystyle \propto \!\!\int \!\!\mathrm{d} k ' \!\exp \!\left[\!- \frac{{k'}^{2}}{2 \!/\! \sigma^{2}} \!\right] \!\exp \!\left[\!- \frac{(\Delta k_{r} \!-\! k')^{2}}{2 \!/\! \sigma^{2}} \!\right] \!e^{-i ( R_{0}^{p} k' + R_{0}^{q} [\Delta k_{r} - k' ] ) } \, , \nonumber
\end{align}
where we used the property that the Fourier transform of a product is given by the convolution of the Fourier transforms and that the Fourier transform of a Gaussian of spread $\sigma$ is a Gaussian of spread ${ 1 / \sigma }$. In  expression~\eqref{shape_integration_WKB}, we note that if ${ |k' | \gg  1 / \sigma }$ or ${ | \Delta k_{r} \!-\! k' | \gg 1 / \sigma }$, then the product of the two terms can be considered to be negligible. We will therefore suppose that  one of the conditions
\begin{equation}
\Delta k_{r} \gg \frac{1}{\sigma} \, \, \, \text{or} \, \, \, \Delta k_{r} = 0 \, ,
\label{assumption_Deltak_sigma}
\end{equation}
holds.
Under this assumption, the term is non-zero only for ${k^{p}_{r} = k^{q}_r}$. Finally, it remains to prove that non zero terms are only obtained when ${ R^{p}_{0} = R^{q}_{0} }$. The peaks of the two Gaussians in equation~\eqref{calcul_condition_orthogonality_WKB} can be considered as sharp and separated if ${ \Delta R_0 = R^{p}_{0} \!-\! R^{q}_{0} }$ satisfies the condition
\begin{equation}
\Delta R_0 \gg \sigma \, \, \, \text{or} \, \, \, \Delta R_{0} = 0 \, .
\label{assumption_DeltaR_sigma}
\end{equation}
To sum up our assumptions so far, we should consider peak-radius $R_{0}$, spread $\sigma$ and radial frequencies $k_r$ such that
\begin{equation}
\Delta R_{0} \gg \sigma \gg \frac{1}{\Delta k_r} \, .
\label{full_assumptions_WKB}
\end{equation}
To these conditions, one must also add the constraints obtained in equations~\eqref{tight_winding_assumption} and~\eqref{assumption_gaussian_WKB_rewriting} via Poisson's equation. With these assumptions, we can ensure that we must necessarily have ${ k^{p}_{\phi} = k^{q}_{\phi} }$, ${ k^{p}_{r} = k^{q}_{r} }$ and ${R^{p}_{0} = R^{q}_0}$, in order to have a non-zero term. The last step is to explicitly calculate the amplitude $\mathcal{A}$ of the basis elements. Indeed, starting from equation~\eqref{calcul_condition_orthogonality_WKB}, we have the condition
\begin{equation}
-  \mathcal{A}^{2} \frac{| k_r |}{G} \frac{1}{\sqrt{\pi} \sigma} \!\int \!\mathrm{d} R \, R \, \exp \left[ - \frac{(R \!-\! R_0)^{2}}{\sigma ^2}\right] = -1 \, ,
\label{calcul_amplitude_orthogonal_I}
\end{equation}
which may be rewritten as
\begin{equation}
1 = \mathcal{A}^{2} \frac{| k_r |}{G}  \frac{R_0}{2} \left[ 1 \!+\! \text{erf} \!\left[\! \frac{R_0}{\sigma} \!\right] \!+\! \frac{1}{\sqrt{\pi}} \frac{\sigma}{R_0} e^{- R_0^2 / \sigma^2}  \right] \, .
\label{calcul_amplitude_orthogonal_II}
\end{equation}
Using the assumptions made in equation~\eqref{full_assumptions_WKB}, we immediately conclude that ${R_{0} / \sigma \gg 1}$, so that ${\text{erf} \left[ R_0 / \sigma \right] \sim 1}$ and ${\sigma / (\sqrt{\pi} R_{0})  \exp [- R_0^2 / \sigma^2] \ll 1}$. We therefore finally obtain the expression of the amplitude of the basis potentials as
\begin{equation}
\mathcal{A} = \sqrt{\frac{G}{| k_{r} | \, R_0}} \, .
\label{expression_amplitude_WKB}
\end{equation}

\subsubsection{Fourier development in angles}
\label{sec:FourierWKB}

The diffusion equation involves terms of the form ${\psi^{(p)}_{\bm{m}} (\bm{J})}$, that we will now evaluate for the WKB basis, equation~\eqref{definition_tightly_spiral}. Using the epicyclic mapping from equation~\eqref{mapping_r_phi_palmer}, we need to estimate
\begin{align}
& \!\!\! \psi_{\bm{m}}^{[ k_{\phi}, k_r , R_0]} (\bm{J}) =  \nonumber
\\
\label{calcul_psi_tight}  & \displaystyle \frac{1}{(2 \pi)^{2}} \!\int \!\!\mathrm{d} \theta_{\phi} \!\int \!\!\mathrm{d} \theta_{R} \, e^{- i m_{\phi} \theta_{\phi} } e^{- i m_{r} \theta_{R}} e^{i k_{\phi} \theta_{\phi}} e^{i k_{r} R_g} \, \times 
\\
&  e^{i [ k_{r} A \cos (\theta_{R}) - k_{\phi} \frac{2 \Omega}{\kappa} \frac{A}{R_{g} } \sin (\theta_R) ] }  \mathcal{A} \,\mathcal{B}_{R_{0}} (R_{g} \!+\! A \cos (\theta_{R})) \, . \nonumber
\end{align}
The integration on $\theta_{\phi}$ is straightforward and is equal to ${ 2 \pi \delta^{m_{\phi}}_{ k_{\phi}} }$. Looking at the dependence in $\theta_{R}$ within the complex exponential, we can write
\begin{equation}
k_{r} A \cos (\theta_{R}) \!-\! k_{\phi} \frac{2 \Omega}{\kappa} \frac{A}{R_{g} } \sin (\theta_R) = H_{k_{\phi}} \!(k_{r}) \sin (\theta_{R} \!+\! \theta^{0}_{R}) \, ,
\label{simplification_calcul_psi_tight} \nonumber
\end{equation} 
where we have defined

\begin{equation}
H_{k_{\phi}} \!(k_r) \!=\! A \sqrt{\! k_{r}^{2} \!+\! k_{\phi}^{2} \!\left[\! \frac{2 \Omega }{\kappa R_{g}} \!\right]^{2} } \; ; \; \theta^{0}_{R} \!=\! \tan^{-1} \!\!\left[\! - \frac{\kappa}{2 \Omega} \frac{k_{r} R_{g}}{k_{\phi}} \!\right] \, .
\label{definition_H_theta_0_tight}
\end{equation}
Thanks to our WKB assumptions, an approximation of the amplitude term $H_{k_{\phi}} \!(k_{r})$ and the phase-shift $\theta_{0}^{R}$ is possible. Indeed, both of these terms involve an expression of the form ${2 k_{\phi} \!\times\! \Omega / \kappa \!\times\! 1 / (k_{r} R_{g})}$. Yet, we made the assumption that ${k_{r} R_{g} \gg 1}$. Moreover, we know that for typical galaxies ${1/2 \leq \Omega / \kappa \leq 1}$ \citep{BinneyTremaine2008}. Assuming that $k_{\phi}$ is of the order of  unity, we obtain the approximations
\begin{equation}
H_{k_{\phi}} \!(k_{r}) \simeq A \, | k_{r} |  \simeq \sqrt{\frac{2 J_{r}}{\kappa}} \, | k_{r} | \;\;\; ; \;\;\; \theta_{R}^{0} \simeq -\frac{\pi}{2} \, .
\label{approximation_amplitude_phase_shift}
\end{equation}
We also supposed that the radial oscillations are small, so that the epicyclic amplitude satisfies ${ A \ll R_g }$. Thanks to this assumption, we may replace ${ \mathcal{B}_{R_{0}} (R_g \!+\! A \cos (\theta_R)) }$ by ${ \mathcal{B}_{R_{0}} (R_g) }$, keeping the dependence on $A$ only in the complex exponential. This is a crucial step to be able to integrate explicitly on $ \theta_R$. We also introduce the Bessel functions of the first kind $\mathcal{J}_{\ell}$ which satisfy the property
\begin{equation}
e^{i H_{k_{\phi}} \! (k_{r}) \sin (\theta_R + \theta^{0}_{R})} = \sum\limits_{\ell \in \mathbb{Z}} \mathcal{J}_{\ell} [H_{k_{\phi}} \!(k_{r})] \, e^{i \ell (\theta_R + \theta^{0}_R)} \, .
\label{definition_Bessel_functions}
\end{equation}
It is then possible to perform explicitly the integration on $\theta_R$ in equation~\eqref{calcul_psi_tight}, which is equal to $2 \pi \, \delta^{m_{r}}_{ \ell} $,  so that only one Bessel function remains. We finally obtain the expression of the Fourier transform in angles of the basis elements
\begin{equation}
\psi^{\![k_{\phi}, k_{r}, R_0\!]}_{\bm{m}} \!(\!\bm{J}) \!=\! \delta^{k_{\phi}}_{m_{\phi}}  e^{\!i k_{r} R_{g} } \!e^{\!i m_{r} \theta^{0}_{R}} \! \mathcal{A} \, \mathcal{J}_{m_{r}} \![H_{\!m_{\phi}} \!(\!k_r\!)] \, \mathcal{B}_{\!R_{0}} \!(\!R_g\!) . 
\label{calcul_psi_tight_II}
\end{equation}

\subsection{Estimation of the response matrix}
\label{sec:estimationresponsematrix}

Using the explicit WKB potential basis introduced in equation~\eqref{definition_tightly_spiral}, one can now estimate the matrix response from  expression~\eqref{Fourier_M}. The approximation obtained in equation~\eqref{approximation_amplitude_phase_shift} allows us to simplify the phase-shift terms, so that equation~\eqref{Fourier_M} becomes
\begin{equation}
\begin{aligned}
& \!\!\!\!\widehat{\mathbf{M}}_{\left[ k^{p}_{\phi}, k^{p}_{r} ,R^{p}_0 \right] , \left[k^{q}_{\phi},  k^{q}_{r} ,R^{q}_0 \right]} (\omega)  = 
\\
& (2 \pi)^{2} \!\sum\limits_{\bm{m}} \!\!\int \!\!\mathrm{d}^{2} \!\bm{J} \, \frac{\bm{m} \!\cdot\! \partial F_{0} \!/\! \partial \bm{J}}{\omega \!-\! \bm{m} \!\cdot\! \bm{\Omega}} \delta^{k^{p}_{\phi}}_{m_{\phi}} \delta^{k^{q}_{\phi}}_{m_{\phi}} e^{i R_{g} \left[ k^{q}_{r} - k^{p}_{r} \right]} \mathcal{A}_{p} \mathcal{A}_{q} \,  \times
\\
& \mathcal{J}_{m_{r}} (H_{m_{\phi}} \!(k^{p}_{r})) \, \mathcal{J}_{m_{r}} (H_{m_{\phi}} \!(k^{q}_{r})) \, \mathcal{B}_{R_{0}^{p}} (R_g) \,\mathcal{B}_{R_{0}^{q}} (R_g) \, .
\end{aligned}
\label{calcul_M_tight}
\end{equation}
One should note that this expression is similar to  equation~\eqref{calcul_condition_orthogonality_WKB}. Indeed, using the assumptions from equation~\eqref{full_assumptions_WKB}, we are able to ensure that only the diagonal coefficients of the response matrix are different from 0. First of all, the azimuthal Kronecker symbols impose that $k_{\phi}^{p} = k_{\phi}^{q}$. There is however a slight complication in the calculation because of the presence of additional terms depending on $R_{g}$. In order to sketch the proof of this statement, we introduce the function 
\begin{align}
\label{definition_h_sketch_diagonal_matrix}  \!\!\!h (\!R_{g}\!) \!=\! & \left|\! \frac{\mathrm{d} J_{\phi}}{\mathrm{d} R_{g}} \!\! \right| \!\frac{\bm{m} \!\cdot\! \partial F_{0} \!/\! \partial \bm{J}}{\omega \!-\! \bm{m} \!\cdot\! \bm{\Omega}} \!\mathcal{A}_{p} \! \mathcal{A}_{q} \mathcal{J}_{m_{r}} \!\!\!\left[ \!\!\sqrt{\!\tfrac{2 J_{r}}{\kappa}} k_{r}^{p} \!\right] \!\!\mathcal{J}_{m_{r}} \!\!\!\left[ \!\!\sqrt{\!\tfrac{2 J_{r}}{\kappa}} k_{r}^{q} \!\right] . 
\end{align}
This function captures all the additional $R_{g}$ dependence appearing in equation~\eqref{calcul_M_tight}. Using the change of variables ${ J_{\phi} \mapsto R_{g} }$, the integral on $J_{\phi}$ which has to be evaluated in equation~\eqref{calcul_M_tight} takes the form
\begin{equation} \hskip -0.35cm
\; \int \!\!\mathrm{d} R_{g} h (R_{g}) \, e^{i R_{g} [ k_{r}^{q} - k_{r}^{p} ]} \!\exp \!\left[\! - \frac{(R_{g} \!\!-\!\! R_{0}^{p})^{2}}{2 \sigma^{2}} 
\!-\! \frac{(R_{g} \!\!-\!\! R_{0}^{q})^{2}}{2 \sigma^{2}} \!\right] .
\label{sketch_diagonal_response_matrix_I}
\end{equation}
Using the assumption from equation~\eqref{assumption_DeltaR_sigma} relative to the possible values of ${ \Delta R_{0} }$ in the WKB basis, one can note that the product of the two Gaussians in $R_{g}$ imposes ${ R_{0}^{p} = R_{0}^{q} }$ in order to have a non-zero contribution. The previous expression then becomes
\begin{equation}
\eqref{sketch_diagonal_response_matrix_I} \propto \int \!\mathrm{d} R_{g} \, h (R_{g}) \, e^{i R_{g} [ k_{r}^{q} - k_{r}^{p} ]} \, \exp \!\left[ - \frac{(R_{g} \!-\! R_{0})^{2}}{\sigma^{2}}\right] \, .
\label{sketch_diagonal_response_matrix_II}
\end{equation}
Using the same argument as in equation~\eqref{shape_integration_WKB}, we can rewrite the Fourier transform in $R$ as the convolution of two radial Fourier transforms so that it becomes
\begin{equation}
\eqref{sketch_diagonal_response_matrix_I} \propto \int \mathrm{d} k' \, \mathcal{F} [h] (k') \, \exp \!\left[ - \frac{(\Delta k_{r} \!-\! k')^{2}}{4 \!/\! \sigma^{2} } \right] \, .
\label{sketch_diagonal_response_matrix_III}
\end{equation}
We now use equation~\eqref{assumption_Deltak_sigma} relative to the possible values of ${ \Delta k_{r} }$ in the WKB basis. If we suppose that ${\Delta k_{r} \neq 0}$, the width of the Gaussian from equation~\eqref{sketch_diagonal_response_matrix_III} imposes that the integration will only \textit{probe} the contribution of $\mathcal{F}[h]$ in the neigborhood of ${k' \sim \Delta k_{r} \gg 1 / \sigma}$. We assume that the radial Fourier transform of the function $h$ is such that it is mainly \textit{focused} in the frequency region ${|k| \lesssim 1 / \sigma}$, meaning that for a typical galactic disc, the main frequencies of radial variations of $h$ are inferior to ${1 / \sigma}$. Under this assumption of slow radial variation within the disc, one can see that non-zero contributions can only be obtained for ${\Delta k_{r} = k_{r}^{p} \!-\! k_{r}^{q} = 0}$. As a consequence, we have shown that within the WKB approximation the response matrix is diagonal. For these diagonal coefficients, it only remains to evaluate explicitly the integrals over $J_{\phi}$ and $J_{r}$ in order to obtain the expression of the response matrix eigenvalues. This calculation is presented in Appendix~\ref{sec:responsematrixcoefficients}. Within the assumption that the galactic disc is tepid, the eigenvalues of the response matrix take the form
\begin{equation}
\widehat{\mathbf{M}}_{\left[  k^{p}_{\phi}, k^{p}_{r} ,R_0 \right] , \left[ k^{q}_{\phi}, k^{q}_{r} ,R_0 \right]} (\omega) \!=\! \delta_{k^{p}_r}^{k^{q}_r}  \delta_{k^{p}_{\phi}}^{k^{q}_{\phi}} \frac{2 \pi G \, \Sigma \, | k_r |}{\kappa^2 (1 \!-\! s^2)}  \mathcal{F} (s, \chi) \, ,
\label{calcul_M_tight_tepid}
\end{equation}
where $\chi$ and $s$ are respectively defined in equations~\eqref{definition_chi} and~\eqref{definition_s} and $F (s , \chi)$ is the reduction factor introduced in equation~\eqref{definition_F_G}. This amplification eigenvalue is in full agreement with the seminal works from~\cite{Kalnajs1965} and~\cite{Lin1966}, which independently derived the WKB dispersion relation for stellar discs.\footnote{For nice introductions to the WKB dispersion relation in stellar discs, see section $6.2.2$ of~\cite{BinneyTremaine2008} and section $1.4.2$ of~\cite{BinneySecular2013}.}

\subsection{Estimation of the diffusion coefficients}
\label{sec:estimationdiffusioncoefficients}

The expression~\eqref{expression_Re_Dm_averaged} of the diffusion coefficients shows that the diffusion coefficients require the evaluation of $[ \mathbf{I} \!-\! \widehat{\mathbf{M}} ]^{-1}$. In order to simplify the notations, we will denote our potential basis with only one index so that
\begin{equation}
\psi^{(p)} = \psi^{[k^{p}_{\phi},k^{p}_r ,R^{p}_{0}]}\, .
\label{redifinition_psi}
\end{equation}
Equation~\eqref{calcul_M_tight_tepid} shows that the response matrix is diagonal in the WKB approximation. We therefore introduce the eigenvalues of $\widehat{\mathbf{M}}$ as
\begin{equation}
\lambda_{p} \equiv \widehat{\mathbf{M}}_{p p} \, .
\label{definition_eigenvalues_M}
\end{equation}
The matrix $[ \mathbf{I} \!-\! \widehat{\mathbf{M}}]^{-1}$ is then diagonal and reads
\begin{equation}
[ \mathbf{I} \!-\! \widehat{\mathbf{M}} ]^{-1}_{p q} = \delta_{p}^{q} \, \frac{1}{1 \!-\! \lambda_{p}} \, .
\label{calcul_I-M}
\end{equation}
Thanks to these diagonal coefficients, the expression of the diffusion coefficients from equation~\eqref{expression_Re_Dm_averaged} becomes
\begin{equation}
D_{\bm{m}} (\bm{J}) = \frac{1}{2} \sum\limits_{p,q} \psi^{(p)}_{\bm{m}} \, \psi^{(q) *}_{\bm{m}} \frac{1}{1 \!-\! \lambda_{p}} \frac{1}{1 \!-\! \lambda_{q}} \, \widehat{\mathbf{C}}_{pq} (\bm{m} \!\cdot\! \bm{\Omega}) \, .
\label{evaluation_diffusion_coefficients_WKB}
\end{equation}
At this stage, we use the property from equation~\eqref{autocorrelation_Fourier_II} to rewrite $\widehat{\mathbf{C}}_{pq}$ as a function of the basis coefficients $\widehat{b}_{p}$ and $\widehat{b}_{q}^{*}$. Remembering that the basis elements $\psi_{\bm{m}}^{(p)}$ and the matrix eigenvalues $\lambda_{p}$ do not change from one realization to another, one can rewrite equation~\eqref{evaluation_diffusion_coefficients_WKB} under the form
\begin{align}
D_{\bm{m}} (\bm{J}) = & \,\bigg< \frac{1}{2 \pi} \!\int \!\! \mathrm{d} \omega ' \frac{1}{2} \sum\limits_{p , q} \psi^{(p)}_{\bm{m}} (\bm{J} )\, \psi^{(q) \, *}_{\bm{m}} (\bm{J}) \, \times \nonumber
\\
& \;\;\;\label{evaluation_diffusion_coefficients_WKB_II} \frac{1}{1 \!-\! \lambda_{p}} \frac{1}{1 \!-\! \lambda_{q}} \widehat{b}_{p} (\bm{m} \!\cdot\! \bm{\Omega}) \, \widehat{b}_{q}^{*} (\omega ') \bigg> \, .
\end{align}
It is important here to note that the eigenvalues $\lambda_{p}$, $\lambda_{q}$ and the basis coefficient $\widehat{b}_{p}$ are both evaluated at the intrinsic frequency ${\bm{m} \!\cdot\! \bm{\Omega}}$, whereas $\widehat{b}_{q}^{*}$ has to be evaluated at the \textit{integrated} frequency $\omega'$. In the upcoming calculations, in order to shorten the notations, when obvious, the frequencies of evaluation will not be written. Using the expressions of the basis elements in the WKB approximation from equation~\eqref{calcul_psi_tight_II}, we can write
\begin{align}
& D_{\bm{m}} (\bm{J}) = \bigg< \frac{1}{2 \pi} \! \int \!\! \mathrm{d} \omega '  \!\!\!\!\!\!\! \sum\limits_{k_{r}^{p} , k_{r}^{q} , R_{0}^{p} , R_{0}^{q}} \! \frac{1}{2} \frac{G}{\sqrt{R_{0}^{p} R_{0}^{q}}} \frac{1}{\sqrt{|k_{r}^{p} k_{r}^{q}|}} \, \times \nonumber
\\
& \mathcal{J}_{m_r} \!\!\left[\! \sqrt{\!\tfrac{2 J_{r}}{\kappa}} k_{r}^{p} \!\right] \mathcal{J}_{m_{r}} \!\!\left[\! \sqrt{\!\tfrac{2 J_{r}}{\kappa}} k_{r}^{q} \!\right] e^{i R_{g} (k_{r}^{p} - k_{r}^{q})} \frac{1}{1 \!-\! \lambda_{p}} \frac{1}{1 \!-\! \lambda_{q}} \, \times \nonumber
\\
\label{calcul_diffusion_coefficients_WKB} &  \frac{1}{\sqrt{\pi \sigma^{2}}} \exp \!\left[ - \frac{(R_{g} \!-\! R_{0}^{p})^{2}}{2 \sigma^{2}} \right]  \exp \!\left[ - \frac{(R_{g} \!-\! R_{0}^{q})^{2}}{2 \sigma^{2}} \right] \widehat{b}_{p} \, \widehat{b}_{q}^{*} \bigg> \, ,
\end{align}
where we already got rid of the sum over $k_{\phi}^{p}$ and $k_{\phi}^{q}$, since the Fourier transform of the WKB basis elements from equation~\eqref{calcul_psi_tight_II} imposes to have 
\begin{equation}
m_{\phi} = k_{\phi}^{p} = k_{\phi}^{q} \, .
\label{equality_coefficients_mphi}
\end{equation}
In the expression~\eqref{calcul_diffusion_coefficients_WKB}, we also neglected the phase terms in ${m_{r} \theta_{0}^{R}}$ and simplified the value at which the Bessel functions have to be evaluated using the approximation introduced in equation~\eqref{approximation_amplitude_phase_shift}.

 In order to obtain an expression independent from the choice of the basis (i.e. the precise value of $\sigma$), we will now replace the coefficients $\widehat{b}_{p}$ by their expressions in terms of the \textit{true} exterior potential function $\psi^{e}$, which is completely independent of the choice of the basis. As the potential basis in the WKB approximation is bi-orthogonal, the temporal Fourier transform of the basis coefficients is given by
\begin{equation}
\widehat{b}_{p} (\omega) = - \!\int \!\!\mathrm{d}^{2} \bm{x} \left[ \Sigma^{(p)} (\bm{x}) \right]^{*} \widehat{\psi^{e}} (\bm{x},\omega) \, ,
\label{calcul_autocorrelation_I}
\end{equation}
where the hat $\widehat{ \, \cdot \, }$ corresponds to the temporal Fourier transform defined in equation~\eqref{definition_Fourier_temporal}. Using the expression of the surface density basis from equation~\eqref{orthogonality_relation_WKB_complex}, we obtain
\begin{equation}
\widehat{b}_{p} =  \!\int \!\!\mathrm{d} R \, R  \!\int \!\!\mathrm{d} \phi \, \frac{|k_r^{p}|}{2 \pi G} \, \mathcal{A} \, e^{- i [ k_r^{p} R + k_{\phi}^{p} \phi ]} \, \mathcal{B}_{R_0^{p}} (R) \, \widehat{\psi^{e}} (R, \phi) \, .
\label{calcul_autocorrelation_II} \nonumber
\end{equation}
The integration on $\phi$ is straightforward and leads to a term equal to $2 \pi \, \widehat{\psi^{e}}_{\!k_{\phi}^{p}} (R) $, where the presence of the index $k_{\phi}^{p}$ corresponds to the Fourier transform with respect to the physical angle $\phi$, using the same conventions as in equation~\eqref{definition_Fourier}. We may now write
\begin{equation}
\! \widehat{b}_{p}  \!= \!\sqrt{\!\!\frac{| k_r^{p} | }{G \! R_{0}^{p} }}  \frac{1}{(\!\pi \sigma^2\!)^{1\!/\!4}} \!\!\!\int \!\!\!\mathrm{d} R R \exp \!\!\left[\! - \frac{(\!R \!\!-\!\! R_{0}^{p}\!)^{2}}{2  \sigma^{2}} \!\right] \! e^{\!- i R k_r^{p}} \widehat {\psi^{e}}_{\!\!\!k_{\phi}^{p}} \!(\!R) \, .
\label{calcul_autocorrelation_III}
\end{equation}
This integration should be interpreted as the radial Fourier transform at the frequency $k_{r}^{p}$ of the exterior potential in the region close to $R_{0}^{p}$. Since the integrand contains a Gaussian in $R$ of spread $\sigma$, we may take the term in $R$ out of the integral and consider it to be equal to $R_0^{p}$. We now define the local Fourier transform of the exterior potential on a restricted region of radius \citep{Gabor1946} as
\begin{equation}
\hskip -0.1cm
\!\widehat{\psi^{e}}_{\!\!k_{\phi},k_{r}} \![R_{0}] \!=\! \frac{1}{2 \pi} \!\!\!\int \!\!\!\mathrm{d} R \, \widehat{\psi^{e}}_{\!\!k_{\phi}} \!(\!R) \exp \!\!\left[\! - \frac{(\!R \!\!-\!\! R_{0}\!)^{2}}{2 \sigma^{2}} \!\right] \!e^{\!- i (R \!-\! R_{0}) k_{r}}  \!.
\label{definition_local_Fourier}
\end{equation}
This definition is motivated by the fact that if we consider the case of an uniform perturbation $\psi^{e} \!=\! 1$, then its local Fourier transform is independent of $R_{0}$. One may note that this definition is not independent of the decoupling scale $\sigma$, but as we will see later on, it is the relevant quantity in order to obtain diffusion coefficients independent of this \textit{ad hoc} parameter. Thanks to this definition, the basis coefficients from equation~\eqref{calcul_autocorrelation_III} become
\begin{equation}
\widehat{b}_{p} = \sqrt{\frac{| k_{r}^{p} | R_{0}^{p}}{G}} \frac{2 \pi}{(\pi \sigma^{2})^{1\!/\!4}} \, e^{- i R_{0}^{p} k _{r}^{p}} \, \widehat{\psi^{e}}_{\! k_{\phi}^{p} , k_{r}^{p}} [R_{0}^{p}] \, .
\label{calcul_autocorrelation_new}
\end{equation}
We recall the notation used for the exterior potential in the previous expression.  The index $k_{\phi}^{p}$ corresponds to the azimuthal Fourier transform with respect to the physical angle $\phi$, and the index $k_{r}^{p}$ corresponds to the \textit{local} radial Fourier transform with respect to the physical radius $R$ in the neighborhood of $R_{0}^{p}$, as defined in equation~\eqref{definition_local_Fourier}. The diffusion coefficients from equation~\eqref{calcul_diffusion_coefficients_WKB} are then given by
\begin{align}
 D_{\bm{m}} (\bm{J}) = & \, \bigg< \!\frac{1}{2 \pi} \!\! \int \!\! \mathrm{d} \omega '  \!\!\!\!\!\!\!\!\!\!\sum\limits_{k_{r}^{p} , k_{r}^{q} , R_{0}^{p} , R_{0}^{q}} \!\!\!\!\!\!\!\!\!\! \mathcal{J}_{m_r} \!\!\left[\! \sqrt{\!\tfrac{2 J_{r}}{\kappa}} k_{r}^{p} \!\right] \!\mathcal{J}_{m_{r}} \!\!\left[\! \sqrt{\!\tfrac{2 J_{r}}{\kappa}} k_{r}^{q} \!\right] \times \nonumber
\\
& e^{i (R_{g} - R_{0}^{p}) k_{r}^{p}} \, e^{- i (R_{g} - R_{0}^{q}) k_{r}^{q}} \frac{1}{1 \!-\! \lambda_{p}} \frac{1}{1 \!-\! \lambda_{q}} \, \times \nonumber
\\
 & \frac{2 \pi}{\sigma^{2}} \exp \!\left[ - \frac{(R_{g} \!-\! R_{0}^{p})^{2}}{2 \sigma^{2}} \right] \exp \!\left[ - \frac{(R_{g} \!-\! R_{0}^{q})^{2}}{2 \sigma^{2}} \right] \, \times \nonumber
 \\
 \label{estimation_diffusion_coefficients_final} & \widehat{\psi^{e}}_{\! m_{\phi} , k_{r}^{p}} [R_{0}^{p}] \, \widehat{\psi^{e}}_{\! m_{\phi} , k_{r}^{q}}^{\, *} [R_{0}^{q}] \bigg> \, .
\end{align}
One can note that the gravitational constant $G$ has disappeared, since the dependence on the strength of the gravity is now \textit{hidden} in the units of $\psi^{e}$. One should also recall that the previous expression has to be evaluated at the resonance frequency, so that ${ \omega = \bm{m} \!\cdot\! \bm{\Omega} }$, except for ${ \widehat{\psi^{e}}_{\!\! m_{\phi} , k_{r}^{q}}^{\, *} [R_{0}^{q}] }$ which has to be evaluated at the frequency ${ \omega' }$. The main step of the simplification is now to replace the discrete sums on the basis index $k_{r}^{p}$, $k_{r}^{q}$, $R_{0}^{p}$ and $R_{0}^{q}$ by continuous integrals. One should indeed  now recall that our potential basis elements are made of three different index. Here $k_{\phi}$ is a discrete index which must necessarily be equal to $m_{\phi}$, so that it is absent from the sums, $k_{r}$ is a continuous index, whose value has to belong to ${] 1 / \sigma \,;\, ...[}$, because of the approximations made in equation~\eqref{full_assumptions_WKB}, and finally $R_{0}$ whose values belong to ${] \sigma \,;\, ... ]}$. We must also comply with the two assumptions~\eqref{assumption_Deltak_sigma} and~\eqref{assumption_DeltaR_sigma} about the distance $\Delta k_{r}$ and $\Delta R_0$ between two consecutive elements of the basis. In order to get rid of the sum over the discrete index, we will use Riemann formula ${\sum f (x) \Delta x \!\simeq\! \int \!\mathrm{d} x \, f(x)}$, with ${\Delta x}$ controlling the distance between two consecutive elements. The dependences with the two radial frequencies $k_{r}^{p}$, $k_{r}^{q}$ and the two radii $R_{0}^{p}$, $R_{0}^{q}$ are such that the sums on the index $p$ can be completely disentangled from the sums on the index $q$. In order to emphasize the gist of the calculation, the diffusion coefficients from equation~\eqref{estimation_diffusion_coefficients_final} may be written under the form
\begin{equation}
D_{\bm{m}} (\bm{J}) = \bigg<\! \frac{1}{2 \pi} \!\! \int \!\!\mathrm{d} \omega '  g (\bm{m} \!\cdot\! \bm{\Omega}) \, g^{*} \!(\omega ') \!\bigg> \, ,
\label{simplification_diffusion_coefficients_discrete}
\end{equation}
where $g (\omega)$ is defined as
\begin{equation}
g (\omega) = 2 \pi \!\!\sum\limits_{k_{r}^{p} , R_{0}^{p}} \!\!g_{s} (k_{r}^{p} , R_{0}^{p} , \omega) \, e^{i (R_{g} - R_{0}^{p}) k_{r}^{p}} \, \mathcal{G} (R_{g} \!-\! R_{0}^{p}) .
\label{definition_g_calculation_diffusion_coefficients}
\end{equation}
In equation~\eqref{definition_g_calculation_diffusion_coefficients}, ${g_{s} (k_{r}^{p} , R_{0}^{p} , \omega)}$ encompasses all the \textit{slow} dependences of the diffusion coefficients with respect to the position $R_{0}$ and the radial frequency $k_{r}$ so that
\begin{equation}
g_{s} (k_{r}^{p} , R_{0}^{p} , \omega) = \mathcal{J}_{m_{r}} \!\!\left[\! \sqrt{\!\tfrac{2 J_{r}}{\kappa}} k_{r}^{p} \!\right] \!\frac{1}{1 \!-\! \lambda_{k_{r}^{p}}} \, \widehat{\psi^{e}}_{\!\! m_{\phi} , k_{r}^{p}} [R_{0}^{p} , \omega] \, ,
\label{definition_gs_calculation_diffusion_coefficients}
\end{equation}
and $\mathcal{G} (R_{g} \!-\! R_{0}^{p})$ is a normalized Gaussian given by
\begin{equation}
\mathcal{G} (R_{g} \!-\! R_{0}^{p}) = \frac{1}{\sqrt{2 \pi \sigma^{2}}} \, \exp \!\left[ - \frac{(R_{g} \!-\! R_{0}^{p})^{2}}{2 \sigma^{2}}\right] \, .
\label{definition_gaussian_simplification_diffusion_coefficients}
\end{equation}
In the discrete sum from equation~\eqref{definition_g_calculation_diffusion_coefficients}, the basis elements are separated by constant step distances $\Delta R_{0}$ and $\Delta k_{r}$. We suppose that generally $k_{r}^{p}$ and $R_{0}^{p}$ are given by
\begin{equation}
\begin{cases}
k_{r}^{p} = n_{k} \Delta k_{r} \, ,
\\
R_{0}^{p} = R_{g} + n_{r} \Delta R_{0} \, ,
\end{cases}
\label{step_dependence_WKB}
\end{equation}
where $n_{k}$ is a strictly positive integer and $n_{r}$ is an integer that can be both positive and negative. One can note in equation~\eqref{definition_g_calculation_diffusion_coefficients} the presence of a rapidly evolving complex exponential which may cancel out the diffusion coefficients if the basis step distances are not chosen carefully. Injecting the dependences from equation~\eqref{step_dependence_WKB} in the complex exponential from equation~\eqref{definition_g_calculation_diffusion_coefficients}, one can see that we have to sum terms of the form
\begin{equation}
\exp \!\Big(\! {i (R_{g} \!-\! (R_{g} \!+\! n_{r} \Delta R_{0})) \, n_{k} \Delta k_{r}} \!\Big) = \exp \!\Big(\! {- i n_{r} n_{k} \Delta R_{0} \Delta k_{r}} \! \Big)\, . \nonumber
\label{shape_complex_exponential}
\end{equation}
As a consequence, since ${ n_{r} n_{k} }$ is an integer, in order to have no contributions from the complex exponential term, one has to choose step distances so that
\begin{equation}
\Delta R_{0} \, \Delta k_{r} = 2 \pi \, .
\label{choice_step_distances}
\end{equation}
This choice of step distances, imposed by the complex exponential term, corresponds to a critical sampling~\citep{Daubechies1990}, which allows us when performing the change to continuous expression in equation~\eqref{definition_g_calculation_diffusion_coefficients} to leave out the complex exponential. This transformation is a subtle stage of the calculation, since we require our step distances $\Delta R_{0}$ and $\Delta k_{r}$ to be simultaneously \textit{large} to comply with the WKB constraints from equation~\eqref{full_assumptions_WKB} and \textit{small} to allow the use of Riemann sum formula. As the radial Gaussian ${\mathcal{G} (R_{g} \!-\! R_{0}^{p})}$ is sufficiently peaked and correctly normalized, one can replace it by ${ \delta_{\rm D} (R_{g} \!-\! R_{0}^{p}) }$. The integration on $R_{0}^{p}$ can then be immediately performed to obtain
\begin{equation}
g (\omega) = \!\int \! \mathrm{d} k_{r}^{p} \, g_{s} (k_{r}^{p} , R_{g} , \omega) \, .
\label{calculation_g_diffusion_coefficients}
\end{equation}
Using this result in equation~\eqref{simplification_diffusion_coefficients_discrete}, we finally obtain the expression of the diffusion coefficients as
\begin{align}
D_{\bm{m}} (\bm{J}) \!=\! \bigg< \! & \frac{1}{2 \pi} \!\!\int \!\! \mathrm{d} \omega ' \!\!\!\int \!\!\! \mathrm{d} k_{r}^{p} \mathcal{J}_{m_{r}} \!\!\!\left[\! \sqrt{\!\tfrac{2 J_{r}}{\kappa}} k_{r}^{p} \!\right] \!\!\frac{1}{1 \!-\! \lambda_{k_{r}^{p}}} \widehat{\psi^{e}}_{\!\! m_{\phi} , k_{r}^{p}} [R_{g}] \, \times \nonumber
\\
& \label{simplification_diffusion_coefficients_II} \int \!\!\! \mathrm{d} k_{r}^{q} \mathcal{J}_{m_{r}} \!\!\!\left[\! \sqrt{\!\tfrac{2 J_{r}}{\kappa}} k_{r}^{q} \!\right] \!\!\frac{1}{1 \!-\! \lambda_{k_{r}^{q}}} \widehat{\psi^{e}}_{\!\! m_{\phi} , k_{r}^{q}}^{*} [R_{g}] \bigg>\,.
\end{align}
In this expression, all the radial functions have to be evaluated at the position $R_{g}$ and at the temporal frequency ${\bm{m} \!\cdot\! \bm{\Omega}}$, except for ${\widehat{\psi^{e}}_{\!\! m_{\phi} , k_{r}^{q}}^{*} [R_{g}]}$ which is evaluated at the frequency $\omega'$. The eigenvalues $\lambda_{k_{r}}$ are given by equation~\eqref{calcul_M_tight_tepid} and read
\begin{equation}
\lambda_{k_{r}} [R_{g} , \bm{m} \!\cdot\! \bm{\Omega}]= \frac{2 \pi G \Sigma | k_{r} |}{\kappa^{2} (1 \!-\! s^{2})} \mathcal{F} (s , \chi) \, .
\label{rewriting_eigenvalues}
\end{equation}
 Note that, as requested, in equation~\eqref{simplification_diffusion_coefficients_II}, all the dependencies in $\sigma$ have disappeared, so that the value of these diffusion coefficients is independent of the precise choice of the WKB basis. One can finally introduce the autocorrelation of the external pertubation $\widehat{C_{\psi}}$ as 
\begin{align}
\label{definition_autocorrelation_C_full} \widehat{C_{\psi}} [m_{\phi} , \omega , k_{r}^{p} , & k_{r}^{q} , R_{g}] = 
\\
& \;\;\;\; \frac{1}{2 \pi} \!\!\int \!\!\! \mathrm{d} \omega ' \Big\langle\! \widehat{\psi^{e}}_{\!\!\!\! m_{\phi} , k_{r}^{p}} [R_{g} , \omega] \, \widehat{\psi^{e}}_{\!\!\!\! m_{\phi} , k_{r}^{q}}^{*} [R_{g} , \omega '] \!\Big\rangle \, , \nonumber
\end{align}
so that the expression~\eqref{simplification_diffusion_coefficients_II} of the diffusion coefficients takes the form
\begin{align}
\label{simplification_diffusion_coefficients_III}  D_{\bm{m}} (\bm{J}) = \! & \int \!\!\! \mathrm{d} k_{r}^{p} \mathcal{J}_{m_{r}} \!\!\left[\! \sqrt{\!\tfrac{2 J_{r}}{\kappa}} k_{r}^{p} \!\right] \!\!\frac{1}{1 \!-\! \lambda_{k_{r}^{p}}} \, \times 
\\
& \int \!\!\! \mathrm{d} k_{r}^{q} \mathcal{J}_{m_{r}} \!\!\left[\! \sqrt{\!\tfrac{2 J_{r}}{\kappa}} k_{r}^{q} \!\right] \!\!\frac{1}{1 \!-\! \lambda_{k_{r}^{q}}} \widehat{C_{\psi}} [m_{\phi} , \!\bm{m} \!\cdot\! \bm{\Omega} , \!k_{r}^{p} , \!k_{r}^{q} , \!R_{g}] \, . \nonumber
\end{align}
We may finally asume that the external perturbations are spatially quasi-stationary, so that we have
\begin{align}
\label{stationarity_translation_assumption_text} \big<\! \psi^{e}_{\!m_{\phi}} [R_{1} , t_{1}] \, & \psi^{e *}_{\!m_{\phi}} [R_{2} , t_{2}] \big> \!=\!
\\
& \;\;\;\;\;\;\; \mathcal{C} \!\left[ m_{\phi} , t_{1} \!-\! t_{2} , R_{1} \!-\! R_{2} , (R_{1} \!+\! R_{2}) / 2 \right] \, , \nonumber
\end{align}
where the dependence of $\mathcal{C}$ with respect to ${ (R_{1} \!\!+\!\! R_{2}) / 2}$ is supposed to be weak. As demonstrated in Appendix~\ref{sec:autocorrelationdiagonalization}, one can then show that
\begin{align}
 &   \big<\! \widehat{\psi^{e}}_{\!\! m_{\phi} , k_{r}^{1}} [R_{g} , \omega_{1}] \, \widehat{\psi^{e}}_{\!\! m_{\phi} , k_{r}^{2}}^{*} [R_{g} , \omega_{2}] \big> = \label{diagonalization_autocorrelation}
\\
& \;\;\;\;\;\;\;\;\;\;\;\;\;\;\;\;\;\;\;\;\;\; 2 \pi \, \delta_{\rm D} (\omega_{1} \!\!-\!\! \omega_{2}) \, \delta_{\rm D} (k_{r}^{1} \!\!-\!\! k_{r}^{2}) \, \widehat{\mathcal{C}} \, [m_{\phi} , \omega_{1} , k_{r}^{1} , R_{g}] \, . \nonumber
\end{align}
Using this autocorrelation function \textit{diagonalized} both in $\omega$ and $k_{r}$, the expression of the diffusion coefficients from equation~\eqref{simplification_diffusion_coefficients_II} finally takes the form
\begin{equation}
\label{simplification_diffusion_coefficients_IV} D_{\bm{m}} (\bm{J}) \!= \!\!\int \!\!\mathrm{d} k_{r}    \frac{ \mathcal{J}^2_{m_{r}}\bigg[ \sqrt{\!\tfrac{2 J_{r}}{\kappa}} k_{r} \!\bigg] }{\displaystyle[1 \!-\! \lambda_{k_{r}}]^2}  \,  
 \widehat{\mathcal{C}} \, [m_{\phi} , \bm{m} \!\cdot\! \bm{\Omega} , k_{r} , R_{g}]  \, .
 \end{equation}
Equation~(\ref{simplification_diffusion_coefficients_IV}) is the main  result of this section.
The corresponding anisotropic tensor diffusion coefficient reads
\begin{equation}
\label{simplification_diffusion_coefficients_IVbis} \mathbf{D} \!= \! \sum_{\bm{m}} \! \bm{m} \!\otimes\! \bm{m} \!\!\int \!\!\mathrm{d} k_{r} \,  \frac{\mathcal{J}^2_{m_{r}}\bigg[ \sqrt{\!\tfrac{2 J_{r}}{\kappa}} k_{r} \!\bigg]}{\displaystyle [1 \!-\! \lambda_{k_{r}}]^2 } \, \widehat{\mathcal{C}} \, [m_{\phi} , \bm{m} \!\cdot\! \bm{\Omega} , k_{r} , R_{g}]
  \, . \nonumber
 \end{equation}
One may sometimes simplify further equation~(\ref{simplification_diffusion_coefficients_IV})  when the function ${ k_{r} \mapsto \lambda_{k_{r}} }$ is a sharp function reaching a maximum value ${\lambda_{\rm max} (R_{g} ,\omega= \bm{m} \!\cdot\! \bm{\Omega})}$, for ${k_{r} = k_{\rm max}}(R_{g} ,\omega)$, with a characteristic spread given by ${\Delta k_{\lambda} }$. Under this assumption of so-called \textit{small denominators}, the previous expression of the diffusion coefficients can be approximated as
\begin{equation}\hskip -0.15cm
D_{\bm{m}} (\!\bm{J}) \!=\!   \Delta k_{\lambda}  \! \frac{\! \mathcal{J}^2_{m_{r}} \!\bigg[\! \sqrt{\!\tfrac{2 J_{r}}{\kappa}} k_{\rm max} \!\bigg] }{\displaystyle[1 \!-\! \lambda_{k_{\rm max}}]^2}   
\, \widehat{\mathcal{C}} \, [m_{\phi} , \bm{m} \!\cdot\! \bm{\Omega} , k_{\rm max} , R_{g}].\label{simplification_diffusion_coefficients_V} \hskip -0.15cm
 \end{equation}
One should note here that the autocorrelation of the external perturbation $\mathcal{C}$, which \textit{sources} the diffusion coefficients ${D_{\bm{m}} (\bm{J})}$ depends on four different parameters: the azimuthal wave number $m_{\phi}$, the location in the disc via $R_{g}$, the radial frequency $k_{\rm max}$ of the most amplified tightly-wound spiral at this position, and finally the local intrinsic frequency ${\bm{m} \!\cdot\! \bm{\Omega}}$.

%
%
%

\section{Discussion and conclusion}
\label{sec:conclusion}

Starting from Boltzmann's collisionless equation expressed in angle-actions coordinates and relying on   a timescale decoupling, we derived in equation~\eqref{diffusion_equation_statistical} a diffusion equation describing the long-term evolution of a perturbed self-gravitating collisionless system\footnote{Appendix~\ref{sec:statistics2} also shows how this diffusion equation is obtained via a different route involving Hamilton's equations.}. This general formalism is  appropriate to capture the \textit{nature} of a collisionless system (via its natural frequencies and susceptibility) as well as its \textit{nurture} via the structure of the power-spectrum of the external perturbations. Hence, it yields the ideal framework in which to study the long-term evolution of such system.

When applying this Fokker-Planck diffusion equation to an infinitely thin galactic disc, we used two main approximations. We first assumed the disc to be tepid. Having orbits with small radial oscillations justified the use of the epicyclic approximation, allowing us to explicitly build up in equation~\eqref{mapping_r_phi_palmer} a mapping between the physical coordinates ${ (\bm{x} , \bm{v}) }$ and the angle-actions coordinates ${ (\bm{\theta} , \bm{J}) }$. Another important consequence of the epicyclic development is to allow for a direct determination of the local frequencies of the system $\Omega$ and $\kappa$, as in equation~\eqref{definition_intrinsic_frequencies}. Being able to {\sl localize} the resonances is crucial in this formalism, since the diffusion coefficients from equation~\eqref{diffusion_equation_statistical} show that both the susceptibility of the system via ${ [\mathbf{I} \!-\! \widehat{\mathbf{M}}] }$ and the external perturbing power spectrum via ${ \widehat{\mathbf{C}} }$ have to be evaluated at the intrinsic frequency ${ \bm{m} \!\cdot\! \bm{\Omega} }$. The second approximation  involves  an explicit WKB basis introduced in equation~\eqref{definition_tightly_spiral}. 
It allowed us to obtain in equation~\eqref{calcul_M_tight_tepid} a diagonal response matrix, as if gravity was only \textit{local}. Thanks to the assumption of radial  decoupling, the WKB approximation led to equation~\eqref{simplification_diffusion_coefficients_IV}, a simple quadrature for the diffusion coefficients, with which it is straightforward to identify the physically relevant modes.
Such simplification provides useful  insight into the  physical processes at work, e.g. the relevant resonances, their loci and their relative strengths.

The formalism of secular resonant dressed orbital diffusion and its WKB limit is implemented in the companion paper \citep[][paper II, submitted]{FouvryPichon2014} to recover the formation of resonant ridges in action-space when an isolated stellar Mestel disc \citep{Mestel1963} is left evolving for hundreds of dynamical times. The development of such ridges has been shown to originate from a resonant mono-dimensional diffusion, specifically enhanced in restricted locations in the disc. It captures the respective roles and importances of various parameters of the system. Indeed, paper II illustrates on an example that  the self-gravity of the disc (via the amplification eigenvalues $\lambda$), its susceptibility (via the anistropic diffusion coefficients $D_{\bm{m}} (\bm{J})$), its inhomogeneity (via the gradients ${ \partial F_{0} / \partial \bm{J} }$), its \textit{temperature} (via $\sigma_{r}^{2}$), its physical structure (via the introduction of tapering functions  representing resp. the bulge and the outer edge of the disc), and the detail of the source of perturbations (via the power spectrum of $\psi^{e}$), all contributes non-negligibly to the appearance of resonant ridges. Such features have been observed both in numerical experiments \citep{Sellwood2012} and in the Solar neighborhood \citep{Wielen1977,Dehnen1998,Nordstrom2004,Famaey2005,AumerBinney2009,McMillan2011}.

The WKB assumption can also be used to study the collisional evolution of a self-gravitating  disc containing a finite number of substructures. Indeed, in \citet[][in prep.]{FouvryPichonChavanis2014}, the same \textit{local} WKB approach will be applied to the Lenard-Balescu non-linear equation \citep{Balescu1960,Lenard1960,Weinberg1998,Heyvaerts2010,Chavanis2012}, which accounts for  self-driven orbital secular diffusion of a self-gravitating system induced by an intrinsic shot noise due to the discreteness of the system. Possible cases of applications of this approach are the secular diffusion of giant molecular clouds in galactic disc, the secular migration of planetesimals in proto-planetary discs, or even the long-term evolution of  population of stars within the Galactic center.

For self-gravitating systems which do not take the form of an infinitely thin disc, for which the epicyclic approximation and the WKB assumption may be relevant, the formalism of secular forcing can still be used. The diffusion equation~\eqref{diffusion_equation_statistical} could for instance describe the secular diffusion of dark matter cusps in galactic centers induced by perturbations from stochastic feedback processes originating from the baryonic disc \citep[][in prep.]{Fouvry2014}. Given a detailed characterization of the perturbations induced by e.g. the cosmic environment, one could also study their long-term effects on a typical self-gravitating collisionless galactic disc. Indeed, in the context of the upcoming \textit{GAIA} mission, this externally induced secular evolution is thought to be a compelling approach to describe the radial migration of stars and its impact on the observed metallicity gradients \citep{SellwoodBinney2002,Roskar2008,Schonrich20092,Solway2012,Minchev2013}. It may also be applied to describe the secular diffusion of accretion streams within the Galactic halo.  Finally, an extension of the formalism of Section~\ref{sec:2dcaseWKB} to  discs with a finite thickness might allow  us to understand the process 
of disc thickening.

\subsection*{Acknowledgements}

JBF thanks the {\sc GREAT} program for travel funding and the department of theoretical physics in Oxford for hospitality. 
CP and JBF thank the Institute of Astronomy, Cambridge, for hospitality 
while this investigation was initiated. 
We thank J. Binney and  P.~H.~Chavanis for detailed comments.
This work is partially supported by the Spin(e) grants ANR-13-BS05-0005 of the French {\sl Agence Nationale de la Recherche}
and by the  LABEX Institut Lagrange de Paris (under reference ANR-10-LABX-63) which  is funded by  ANR-11-IDEX-0004-02.

\bibliographystyle{mn2e}
\bibliography{references}

\appendix

\section{Statistical approach via Hamilton's equation}
\label{sec:statistics2}

We now derive the statistical expression~\eqref{diffusion_equation_statistical} of the secular diffusion coefficients using a different method based on Hamilton's equations and inspired from \cite{Binney1988}.
We will indeed quantify the temporal rate of change of the actions, represented by $\dot{\bm{J}}$. The main difference between the following calculation and that made in \cite{Binney1988} is that we  take explicitly into account the self-gravity of the system which leads to the appearance of a self-perturbing potential $\psi^{s}$ triggered by $\psi^{e}$. Starting from the Hamiltonian introduced in equation~\eqref{perturbation}, and using the Fourier development in angles as in equation~\eqref{definition_Fourier}, Hamilton's equations, ${\dot{\bm{\theta}} = \partial H / \partial \bm{J}}$ and ${\dot{\bm{J}} = - \partial H / \partial \bm{\theta}}$, take the form
\begin{equation}
\begin{cases}
\displaystyle \dot{\bm{\theta}} = \bm{\Omega} + \sum\limits_{\bm{m}} e^{i \bm{m} \cdot \bm{\theta}} \, \frac{\partial }{\partial \bm{J}} \left[ \psi^{s}_{\bm{m}} \!+\! \psi^{e}_{\bm{m}} \right] \, ,
\\
\displaystyle \dot{\bm{J}} = - i \sum\limits_{\bm{m}} \bm{m} \, e^{i \bm{m} \cdot \bm{\theta}} \left[ \psi^{s}_{\bm{m}} \!+\! \psi^{e}_{\bm{m}} \right] \, .
\end{cases}
\label{Hamiltons_equations_Binney_Lacey}
\end{equation}
As we aim to describe the \textit{wandering} in action-space of the particles, we introduce a limited development of the change in actions and angles of the form
\begin{equation}
\begin{cases}
\displaystyle \bm{\theta} (t) = \bm{\theta}_{0} + \bm{\Omega} \, t +\Delta \bm{ \theta} (t) \, ,
\\
\displaystyle \bm{J} (t) = \bm{J}_{0} +\Delta \bm{ J} (t) \, .
\end{cases}
\label{definition_Delta1_Binney_Lacey}
\end{equation}
In order to solve this system of coupled differential equations, we will proceed step-by-step, by including gradually the perturbative terms in ${\psi^{e} \!+\! \psi^{s}}$. First of all, one must note that the unperturbed orbits follow the straight-line trajectories ${(\bm{\theta} , \bm{J}) = (\bm{\theta}_{0} \!+\! \bm{\Omega} t , \bm{J}_{0})}$. Then, the first-order term in action ${\Delta} \bm{J}$ is given by 
\begin{equation}
 \Delta \bm{J}  (T) = \int_{0}^{T} \!\!\mathrm{d} t \, \dot{\bm{J}} (t)\,,
 \end{equation} where $\dot{\bm{J}}$ is given by Hamilton's equations~\eqref{Hamiltons_equations_Binney_Lacey}, where all the occurences of $\bm{\theta} (t)$ and $\bm{J} (t)$ are replaced by the expressions obtained for the unperturbed orbits. After a time $T$, the shift in action at first order is therefore given by
\begin{equation}
\Delta\bm{ J} (T) \!=\!- i \!\sum\limits_{\bm{m}} \!\bm{m} \!\!\int_{0}^{T} \!\!\!\!\mathrm{d} t \left[ \psi^{s}_{\bm{m}} (\bm{J}_{0} , t) \!+\! \psi^{e}_{\bm{m}} (\bm{J}_{0} , t) \right] e^{i \bm{m} \cdot \left( \bm{\theta}_{0} + \bm{\Omega} t \right) } \!. \nonumber
\label{calculation_Delta1_Binney_Lacey}
\end{equation}
We introduce the operation of angle-average on the initial phase $\bm{\theta}_{0}$ as
\begin{equation}
\big\{ F \big\}_{\bm{\theta}_{0}} = \frac{1}{(2 \pi)^{d}} \int \! \mathrm{d} \bm{\theta}_{0} \, F (\bm{\theta}_{0} ) \, .
\label{definition_phase_average_Binney_Lacey}
\end{equation}
In order to characterize the \textit{wandering} in action-space, one has to study the behavior of the \textit{square} of the perturbation $\Delta\bm{ J}$. From \cite{Binney1988}, we know the relation between the \textit{wandering} $\Delta\bm{ J}$ in action-space and the diffusion coefficient appearing in the Fokker-Planck equation, which is given by
\begin{equation}
\mathbf{D}_{i j} (\bm{J}_{0}) = \frac{1}{2 \, T} \big\{\! \Delta J_{i} \, \Delta J_{j} \!\big\}_{\bm{\theta}_{0}} \!(T) \, ,
\label{link_Delta_Diffusion_Coefficient_Binney_Lacey}
\end{equation}
where the diffusion equation has been written under the compact form
\begin{equation}
\frac{\partial F}{\partial t} = \sum\limits_{i , j} \frac{\partial }{\partial J_{i}} \!\left[ \mathbf{D}_{i j} (\bm{J}) \, \frac{\partial F}{\partial J_{j}}\right] \, .
\label{Fokker_Planck_Binney_Lacey}
\end{equation}
Taking an average over the initial phases $\bm{\theta}_{0}$, using the fact that ${\psi^{e/s}_{- \bm{m}} = [\psi^{e / s}_{\bm{m}}]^{*}}$ and projecting the result on the biorthogonal potential basis $\psi^{(p)}$, one can write
\begin{align}
\label{squared_Delta1_basis_Binney_Lacey} & \big\{\! \Delta J_{i} \, \Delta J_{j} \!\big\}_{\bm{\theta}_{0}} \!=\! \sum\limits_{\bm{m}} \sum\limits_{p , q} \bm{m}_{i} \, \bm{m}_{j} \psi^{(p)}_{\bm{m}} (\bm{J}_{0}) \psi^{(q) *}_{\bm{m}} (\bm{J}_{0}) \, \times
\\
& \int_{0}^{T} \!\!\!\!\mathrm{d} t_{1} \!\!\int_{0}^{T} \!\!\!\!\mathrm{d} t_{2}  \big[ a_{p} (t_{1}) \!+\! b_{p} (t_{1}) \big] \big[ a_{q}^{*} (t_{2}) \!+\! b_{q}^{*}  (t_{2}) \big] e^{i \bm{m} \cdot \bm{\Omega} (t_{1} - t_{2})} \, . \nonumber
\end{align}
In order to have an expression which only depends on the exterior potential $\psi^{e}$, we use the convolution relation~\eqref{estimation_a_Laplace}, written as an amplification relation, to obtain
\begin{align}
\big\{\! \Delta J_{i} \,  \Delta J_{j} \!\big\}_{\bm{\theta}_{0}} & \!\!\!=\!  \sum\limits_{\bm{m}} \sum\limits_{p , q} \sum\limits_{k , l} \bm{m}_{i} \, \bm{m}_{j} \, \psi^{(p)}_{\bm{m}} (\bm{J}_{0}) \, \psi^{(q) *}_{\bm{m}} (\bm{J}_{0}) \, \times \nonumber
\\
 \label{squared_Delta1_onlyb_Binney_Lacey} & \!\!\!\!\!\!\!\!\!\!\!\!\!\!\!\!\!\!\!\! \int_{0}^{T} \!\!\!\!\mathrm{d} t_{1} \!\int_{0}^{T} \!\!\!\!\mathrm{d} t_{2} \!\int_{0}^{t_{1}} \!\!\!\!\!\! \mathrm{d} \tau_{1} \, \left[ \mathbf{I} \!-\! \mathbf{M} \right]^{-1}_{p k} \!(t_{1} \!-\! \tau_{1}) \, b_{k} (\tau_{1})  \times
 \\
 &  \!\!\!\!\!\!\!\!\!\!\!\!\!\!\!\!\!\!\!\! \int_{0}^{t_{2}} \!\!\!\!\!\! \mathrm{d} \tau_{2} \left[ \left[ \mathbf{I} \!-\! \mathbf{M} \right]^{-1}_{q l} \right]^{*} \!(t_{2} \!-\! \tau_{2}) \, b_{l}^{*} (\tau_{2}) \, e^{i \bm{m} \cdot \bm{\Omega} (t_{1} \!-\! t_{2})} \, . \nonumber
\end{align}
This four dimensional integral is transformed using the change of variables
\begin{equation}
u_{1} = t_{1} \!-\! \tau_{1} \;;\; u_{2} = t_{2} \!-\! \tau_{2} \;;\; v_{1} = \tau_{1} \!+\! \tau_{2} \;;\; v_{2} = \tau_{1} \!-\! \tau_{2} \, .
\label{change_of_variables_Binney_Lacey}
\end{equation}
It is straightforward to check that the Jacobian of this transformation is $2$, so that equation~\eqref{squared_Delta1_onlyb_Binney_Lacey} becomes
\begin{align}
& \big\{\! \Delta J_{i} \, \Delta J_{j} \!\big\}_{\bm{\theta}_{0}} \!\!=\! \sum\limits_{\bm{m}} \sum\limits_{p , q} \sum\limits_{k , l} \bm{m}_{i} \, \bm{m}_{j} \, \psi^{(p)}_{\bm{m}} (\bm{J}_{0}) \, \psi^{(q) *}_{\bm{m}} (\bm{J}_{0}) \, \times \nonumber
\\
 \label{squared_Delta1_onlyb_Binney_Lacey_II} & \frac{1}{2} \int_{0}^{T} \!\!\!\!\mathrm{d} u_{1} \left[ \mathbf{I} \!-\! \mathbf{M} \right]^{-1}_{p k} \!(u_{1}) \, e^{i \bm{m} \cdot \bm{\Omega} \, u_{1}} \, \times
 \\
 & \int_{0}^{T} \!\!\!\!\mathrm{d} u_{2}  \left[ \left[ \mathbf{I} \!-\! \mathbf{M} \right]^{-1}_{q l} \right]^{*}  \!(u_{2}) \, e^{- i \bm{m} \cdot \bm{\Omega} \, u_{2}} \int_{- (T - u_{2})}^{T - u_{1}} \!\!\!\!\!\!\!\! \mathrm{d} v_{2}  \, \times \nonumber
 \\
 & \int_{|v_{2}|}^{2 T - u_{1} - u_{2} - | v_{2} + u_{1} - u_{2} |} \!\!\!\!\!\!\!\!\!\! \mathrm{d} v_{1} \, b_{k} \!\!\left[\frac{v_{1} \!+\! v_{2}}{2} \right]  b_{l}^{*} \!\!\left[ \frac{v_{1} \!-\! v_{2}}{2} \right] e^{i \bm{m} \cdot \bm{\Omega} \, v_{2}} . \nonumber
\end{align}
We introduce as in equation~\eqref{averaged_diffusion_equation}, the operation of ensemble average over many realizations denoted with ${\langle\,.\,\rangle }$. As in equation~\eqref{stationarity_hypothesis}, we assume that the exterior perturbing potential is a stationary random process. One can then perform an ensemble average of the expression~\eqref{squared_Delta1_onlyb_Binney_Lacey_II}, while assuming, as in equation~\eqref{expression_D_IV}, that the response matrix coefficients can be taken out of the ensemble average operation. We obtain
\begin{align}
\big< \big\{\! \Delta J_{i} \, \Delta J_{j} \!\big\}_{\bm{\theta}_{0}}&  \!\big> = \sum\limits_{\bm{m}} \sum\limits_{p , q} \sum\limits_{k , l} \bm{m}_{i} \, \bm{m}_{j} \, \psi^{(p)}_{\bm{m}} (\bm{J}_{0}) \, \psi^{(q) *}_{\bm{m}} (\bm{J}_{0}) \nonumber
\\
& \frac{1}{2} \int_{0}^{T} \!\!\!\!\mathrm{d} u_{1} \left[ \mathbf{I} \!-\! \mathbf{M} \right]^{-1}_{p k} \!(u_{1}) \, e^{i \bm{m} \cdot \bm{\Omega} \, u_{1}} \, \times \nonumber
\\
& \int_{0}^{T} \!\!\!\!\mathrm{d} u_{2} \left[ \left[ \mathbf{I} \!-\! \mathbf{M} \right]^{-1}_{q l} \right]^{*}  \!(u_{2}) \, e^{- i \bm{m} \cdot \bm{\Omega} \, u_{2}} \, \times \nonumber
\\
\label{squared_Delta1_onlyb_Binney_Lacey_III} & \int_{- (T - u_{2})}^{T - u_{1}} \!\!\!\! \mathrm{d} v_{2}  \, \mathbf{C}_{k l} (v_{2}) \, e^{i \bm{m} \cdot \bm{\Omega} \, v_{2}} \, \times
\\
& \;(2 T \!-\! u_{1} \!-\! u_{2} \!-\! |v_{2} \!+\! u_{1} \!-\! u_{2} | \!-\! |v_{2}|) \, . \nonumber
\end{align}
The next important step of the calculation is to compare $T$ with the various autocorrelation times of the system. The first one is $T_{\rm corr}^{\psi}$ describing the typical autocorrelation time of the realizations of the external perturbations. Two values of the potential perturbations separated by a time larger than $T_{\rm corr}^{\psi}$ can be considered as independent. The second autocorrelation timescale is $T_{\rm corr}^{\mathrm{M}}$, which describes the typical autocorrelation time of the response matrix $\mathbf{M}$ and could be called the \textit{look-back} time. From the expression used in equation~\eqref{squared_Delta1_onlyb_Binney_Lacey}, one can note that the values of the self-response coefficients are obtained via a non-Markovian mechanism, where the past values are amplified thanks to the response matrix. However, the self-gravitating system can not have an \textit{infinite memory}, so that only the sufficiently recent past values should play a role in this amplification. As a consequence, during the amplification process, only the past behavior for a time interval of the order of $T_{\rm corr}^{\mathrm{M}}$ is relevant and amplified, so that $T_{\rm corr}^{\mathrm{M}}$ represents the depth with which the self-response mechanism can probe past values. We finally suppose that the time $T$ for which the wandering in phase-space is studied satisfies the comparison relations
\begin{equation}
T \gg T_{\rm corr}^{\Phi} \;\;\; ; \;\;\; T \gg T_{\rm corr}^{\mathrm{M}} \, .
\label{relation_DeltaT_autocorrelations}
\end{equation}
As a consequence, the integration boundaries appearing in~\eqref{squared_Delta1_onlyb_Binney_Lacey_III} become
\begin{align}
\big< \big\{\! \Delta J_{i} \, \Delta J_{j} \!\big\}_{\bm{\theta}_{0}}&  \!\big> = \sum\limits_{\bm{m}} \sum\limits_{p , q} \sum\limits_{k , l} \bm{m}_{i} \, \bm{m}_{j} \, \psi^{(p)}_{\bm{m}} (\bm{J}_{0}) \, \psi^{(q) *}_{\bm{m}} (\bm{J}_{0}) \nonumber
\\
& \frac{1}{2} \int_{0}^{T_{\rm corr}^{\mathrm{M}}} \!\!\!\!\!\!\mathrm{d} u_{1} \left[ \mathbf{I} \!-\! \mathbf{M} \right]^{-1}_{p k} \!(u_{1}) \, e^{i \bm{m} \cdot \bm{\Omega} \, u_{1}} \, \times \nonumber
\\
& \int_{0}^{T_{\rm corr}^{\mathrm{M}}} \!\!\!\!\!\!\mathrm{d} u_{2} \left[ \left[ \mathbf{I} \!-\! \mathbf{M} \right]^{-1}_{q l} \right]^{*}  \!(u_{2}) \, e^{- i \bm{m} \cdot \bm{\Omega} \, u_{2}} \, \times \nonumber
\\
\label{squared_Delta1_onlyb_Binney_Lacey_IV} & \int_{- T_{\rm corr}^{\psi}}^{T_{\rm corr}^{\psi}} \!\!\!\!\!\! \mathrm{d} v_{2}  \, \mathbf{C}_{k l} (v_{2}) \, e^{i \bm{m} \cdot \bm{\Omega} \, v_{2}} \, \times
\\
& \;(2 T \!-\! u_{1} \!-\! u_{2} \!-\! |v_{2} \!+\! u_{1} \!-\! u_{2} | \!-\! |v_{2}|) \, . \nonumber
\end{align}
Thanks to the assumptions~\eqref{relation_DeltaT_autocorrelations}, one can see that the last term of equation~\eqref{squared_Delta1_onlyb_Binney_Lacey_III} can be approximated by ${ 2  T }$. The remaining integrations can then be seen as truncated temporal Fourier transforms, so that equation~\eqref{squared_Delta1_onlyb_Binney_Lacey_III} becomes
\begin{align}
& \big< \big\{\! \Delta J_{i} \, \Delta J_{j} \!\big\}_{\bm{\theta}_{0}} \!\big> =  \nonumber
\\
& \;\;\;\;\;\;\;\;\;\;\label{squared_Delta1_onlyb_Binney_Lacey_V} T \sum\limits_{\bm{m}} \sum\limits_{p , q} \sum\limits_{k , l} \bm{m}_{i} \, \bm{m}_{j} \, \psi^{(p)}_{\bm{m}} (\bm{J}_{0}) \, \psi^{(q) *}_{\bm{m}} (\bm{J}_{0}) \, \times 
\\
& \;\;\;\;\;\;\;\;\;\; \big[ \mathbf{I} \!-\! \widehat{\mathbf{M}} \big]^{-1}_{p k}  ( \bm{m} \!\cdot\! \bm{\Omega}) \left[\! \big[ \mathbf{I} \!-\! \widehat{\mathbf{M} } \big]^{-1}_{q l}  ( \bm{m} \!\cdot\! \bm{\Omega}) \!\right]^{*}  \!\widehat{\mathbf{C}}_{k l} ( \bm{m} \!\cdot\! \bm{\Omega}) \, . \nonumber
\end{align}
The last step of the simplification is to recall that equation~\eqref{Fourier_M} guarantees that ${\widehat{\mathbf{M}}^{*} \!= \widehat{\mathbf{M}}^{t}}$, so that using equation~\eqref{link_Delta_Diffusion_Coefficient_Binney_Lacey}, we finally obtain the expression of the diffusion coefficients from equation~\eqref{Fokker_Planck_Binney_Lacey} which read
\begin{align}
 \mathbf{D}_{i j} (\bm{J}_{0}) = & \frac{1}{2} \sum\limits_{\bm{m}} \sum\limits_{p , q} \bm{m}_{i} \, \bm{m}_{j} \, \psi^{(p)}_{\bm{m}} (\bm{J}_{0}) \, \psi^{(q) *}_{\bm{m}} (\bm{J}_{0}) \, \times \nonumber
\\
& \label{diffusion_coefficient_final_Binney_Lacey} \left[ [ \mathbf{I} \!-\! \widehat{\mathbf{M}} ]^{-1} \!\!\cdot \,  \widehat{\mathbf{C}} \cdot [ \mathbf{I} \!-\! \widehat{\mathbf{M}} ]^{-1} \!\right]_{p q} \!\!\!( \bm{m} \!\cdot\! \bm{\Omega}) \,\, .
\end{align}
With this approach, we  recover the same diffusion coefficients as the ones obtained in equation~\eqref{diffusion_equation_statistical} via the quasi-linear approach presented in the main text.

\section{WKB response matrix}
\label{sec:responsematrixcoefficients}

We estimate the value of the diagonal response matrix coefficients introduced in equation~\eqref{calcul_M_tight} within the WKB approximation. Using the definition of $h (R_{g})$ from  equation~\eqref{definition_h_sketch_diagonal_matrix} and the fact that $J_{\phi}$ is an increasing function of $R_g$, the integration on $J_{\phi}$ in equation~\eqref{calcul_M_tight} takes the form
\begin{equation}
\int \mathrm{d} R_g \, h(R_g) \frac{1}{\sqrt{2 \pi (\sigma \!/\! \sqrt{2})^2}} \, \exp \!\left[ - \frac{(R_g \!-\! R_0)^{2}}{2 (\sigma \!/\! \sqrt{2})^{2}}\right] \, .
\label{justification_delta_WKB}
\end{equation}
One should note that in this expression, we have a Gaussian of spread ${\sigma \!/\! \sqrt{2}}$ in $R_g$, correctly normalized in order to have an integral over $R_g$ equal to 1. Assuming that this Gaussian is sufficiently peaked, we may replace it by ${ \delta_{\rm D} (R_g \!-\! R_0) }$, so that the integral on $J_{\phi}$ can be dropped. For a Schwarzschild distribution function as in equation~\eqref{complete_DF_Schwarzschild}, we then obtain
\begin{align}
& \widehat{\mathbf{M}}_{\left[  k_{\phi}, k_{r} , R_0 \right] , \left[ k_{\phi}, k_{r} , R_0 \right]} (\omega)  = \nonumber
\\
& (2 \pi)^{2} \mathcal{A}^{2} \left| \frac{\mathrm{d} J_{\phi} }{\mathrm{d} R_g} \right|_{R_{0}} \!\!\frac{\Omega \Sigma}{\pi \kappa \sigma_{r}^{2}} \sum\limits_{m_r}  \, \frac{1}{\omega \!-\! m_{r} \kappa \!-\! k_{\phi} \Omega} \, \times \nonumber
\\
& \bigg\{ \!\!\left[ - m_r \!\frac{\kappa}{\sigma_r^{2}} \!+\! k_{\phi} \!\frac{\partial }{\partial J_{\phi}} \!\!\left[ \ln \!\left(\! \frac{\Omega \Sigma}{\kappa \sigma_r^{2}} \!\right) \!\right] \right]  \!\int \!\!\mathrm{d} J_{r} \, e^{- \frac{\kappa J_r}{\sigma_r^{2}} } \!\mathcal{J}_{m_{r}}^{2} (H_{k_{\phi}} (k_{r})) \nonumber
\\
 \label{calcul_M_tight_II} & \;\;\;- k_{\phi} \frac{\partial }{\partial J_{\phi}} \! \left[ \frac{\kappa}{\sigma_r^{2}} \right] \int \!\!\mathrm{d} J_{r} \, J_{r} \, e^{- \frac{\kappa J_r}{\sigma_r^{2}} } \, \mathcal{J}_{m_{r}}^{2} (H_{k_{\phi}} (k_{r})) \bigg\} \, .
\end{align}
In order to perform the integration on $J_{r}$, the first step is to notice that the only dependence on $J_{r}$ in the Bessel terms is in ${H_{m_{\phi}} \!(k_{r})}$, through ${A = \sqrt{2 J_{r} / \kappa }}$ . Therefore, we will use the two integration formula (see formula (6.615.1) from \cite{Gradshteyn2007})
\begin{numcases}
\displaystyle \int_{0}^{+ \infty} \!\!\!\!\!\!\mathrm{d} J_{r}  e^{- \alpha J_{r}} \,  \mathcal{J}_{m_{r}}^{2} (\beta \sqrt{J_r} ) = \frac{e^{- \beta^2 \!/ 2 \alpha}}{\alpha} \, \mathcal{I}_{m_{r}} \!\left[\! \frac{\beta^{2}}{2 \alpha} \!\right] \, , \nonumber
\\
\int_{0}^{+ \infty} \!\!\!\!\!\!\mathrm{d} J_{r}  J_{r} \, e^{- \alpha J_{r}} \mathcal{J}_{m_{r}}^{2} (\beta \sqrt{J_r} ) = \label{Integration_Bessel}
\\
\;\;\;\;\frac{e^{- \beta^2 \!/ 2 \alpha}}{\alpha^{2} } \bigg\{\!\! \left[ - \frac{\beta^2}{2 \alpha} \!+\! 1 \!+\! |m_{r}| \right]\! \mathcal{I}_{m_{r}} \!\!\left[\! \frac{\beta^2}{2 \alpha} \!\right] \!+\! \frac{\beta^{2}}{2 \alpha} \mathcal{I}_{|m_{r}|+1} \!\left[\! \frac{\beta^2}{2 \alpha} \!\right] \!\!\bigg\} \, . \nonumber
\end{numcases}
where ${\alpha \!>\! 0}$, ${\beta \!>\! 0}$, and ${m_{r} \in \mathbb{Z} }$. First, let's write out explicitly the dependence of ${ \mathcal{H}_{k_{\phi}} \!(k_r) }$ with $J_r$. From equations~\eqref{definition_Jr} and~\eqref{definition_H_theta_0_tight}, we find that
\begin{equation}
\mathcal{H}_{k_{\phi}} \!(k_r) = \sqrt{J_{r}} \, \beta \, , 
\label{link_H_Jr_tight}
\end{equation}
where $\beta$ is defined as
\begin{equation}
\beta = \sqrt{\frac{2}{\kappa} \!\left[ k_{r}^{2} \!+\! k_{\phi}^{2} \!\left[ \frac{2 \Omega}{\kappa R_g}\right]^{2}\right] } \simeq \sqrt{\frac{2}{\kappa}} \, k_{r} \, .
\label{definition_beta}
\end{equation}
This approximate expression has been obtained using the same approximation as in equation~\eqref{approximation_amplitude_phase_shift}. We also introduce the notation 
\begin{equation}
\chi = \frac{\sigma_{r}^{2}}{\kappa^{2}}  \left[ k_{r}^{2} \!+\! k_{\phi}^{2} \!\left[ \frac{2 \Omega}{\kappa R_g}\right]^{2} \right] \simeq \frac{\sigma_{r}^{2} k_{r}^{2}}{\kappa^{2}} \, .
\label{definition_chi}
\end{equation}
We are now able to compute the integrals on $J_{r}$ from equation~\eqref{calcul_M_tight_II}, to obtain
\begin{align}
\widehat{\mathbf{M}}&_{\left[ k_{\phi}, k_{r} , R_0 \right] , \left[ k_{\phi}, k_{r} , R_0 \right]} (\omega) =  \nonumber
\\
& (2 \pi)^{2} \mathcal{A}^{2} \left| \frac{\mathrm{d} J_{\phi} }{\mathrm{d} R_g} \right|_{R_{0}} \!\!\frac{\Omega \Sigma}{\pi \kappa \sigma_{r}^{2}} \sum\limits_{m_r}  \frac{1}{\omega \!-\! m_{r} \kappa \!-\! k_{\phi} \Omega} \, \, \, \times \nonumber
\\
& \label{calcul_M_tight_III} \bigg\{ e^{-\chi} \, \frac{\sigma_r^{2}}{\kappa} \mathcal{I}_{m_r} [\chi] \left[ - m_r \frac{\kappa}{\sigma_r^{2}} \!+\! k_{\phi} \frac{\partial }{\partial J_{\phi}} \!\left[ \ln \!\left(\! \frac{\Omega \Sigma}{\kappa \sigma_r^{2}} \!\right) \!\right]\right]  
\\
 & - k_{\phi} \frac{\partial }{\partial J_{\phi}} \!\!\left[\! \frac{\kappa}{\sigma_r^{2}} \!\right] \!e^{-\chi} \frac{\sigma_{r}^{4}}{\kappa^2} \big[ (1 \!+\! |m_{r}| \!-\! \chi) \, \mathcal{I}_{m_r} [\chi] \!+\! \chi \mathcal{I}_{|m_r|+1} [\chi] \big] \!\!\bigg\} \, . \nonumber
\end{align}
In order to simplify this expression, we recall that we have the property ${ \mathcal{I}_{- m_r} (\chi) \!=\! \mathcal{I}_{m_r} (\chi) }$. Therefore, in equation~\eqref{calcul_M_tight_III}, we have to study four types of sums on $m_r$, which may be simplified as
\begin{numcases}
\!\!\!\!\displaystyle \sum\limits_{m_{r} \in \mathbb{Z}} \!\frac{m_{r} \, \mathcal{I}_{m_{r}} [\chi]}{\omega' \!-\! m_{r} \kappa} = - \frac{2}{\kappa} \!\!\sum\limits_{m_r = 1}^{+ \infty} \!\!\frac{\mathcal{I}_{m_r} [\chi]}{1 \!-\! [ \omega' / m_r \kappa ]^{2}} \, , \nonumber
\\
\!\!\!\!\displaystyle \sum\limits_{m_{r} \in \mathbb{Z}}  \!\frac{\mathcal{I}_{m_{r}} [\chi]}{\omega' \!-\! m_{r} \kappa}   = \frac{\mathcal{I}_{0} [\chi]}{\omega'} \!+\! \frac{2}{\omega'} \!\!\sum\limits_{m_r = 1}^{+ \infty} \!\!\frac{\mathcal{I}_{m_r} [\chi]}{1 \!-\! [ m_r \kappa / \omega' ]^{2}} \, , \label{simplification_sums_Bessel}
\\
\!\!\!\! \displaystyle \sum\limits_{m_{r} \in \mathbb{Z}} \!\frac{|m_{r}| \, \mathcal{I}_{m_{r}} [\chi]}{\omega' \!-\! m_{r} \kappa} = \frac{2}{\omega'} \!\!\sum\limits_{m_{r} = 1}^{+ \infty} \!\!\frac{m_{r} \, \mathcal{I}_{m_{r}} [\chi] }{1 \!-\! [ m_{r} \kappa / \omega' ]^{2} } \, , 
\\
\!\!\!\! \displaystyle \sum\limits_{m_{r} \in \mathbb{Z}} \!\frac{\mathcal{I}_{|m_{r}| + 1} [\chi]}{\omega' \!-\! m_{r} \kappa} = \frac{\mathcal{I}_{1} [\chi]}{\omega'} \!+\! \frac{2}{\omega'} \!\!\sum\limits_{m_{r} = 1}^{+ \infty} \!\!\frac{\mathcal{I}_{m_{r} +1} [\chi]}{1 \!-\! [ m_{r} \kappa / \omega' ]^{2}} \, , \nonumber
\end{numcases}
where we use ${ \omega' = \omega \!-\! k_{\phi} \Omega_{\phi} }$. We define the dimensionless parameter $s$ as
\begin{equation}
s = \frac{\omega \!-\! k_{\phi} \Omega}{\kappa} \, .
\label{definition_s}
\end{equation}
We also introduce the reduction factor ${ \mathcal{F} (s, \chi) }$ \citep{Kalnajs1965,Lin1966} and similar functions ${ \mathcal{G} (s,\chi) }$ , ${ \mathcal{H} (s,\chi) }$ and ${ \mathcal{I} (s , \chi) }$ defined as
\begin{numcases}{\!\!\!\!\!\!\!\!\!\!\!\!}
\!\!\!\! \displaystyle \mathcal{F} (s,\chi) = 2 \, (1 \!-\! s^{2}) \frac{e^{- \chi}}{\chi} \!\!\sum\limits_{m_r = 1}^{+ \infty} \!\!\frac{\mathcal{I}_{m_r} [\chi]}{ 1 \!-\! \big[ s / m_r \big]^{2}} \, , \nonumber
\\
\!\!\!\! \displaystyle \mathcal{G} (s,\chi) = 2 \, (1 \!-\! s^2) \frac{e^{- \chi}}{\chi} \!\left[ \frac{1}{2} \frac{\mathcal{I}_{0} [ \chi ]}{s} \!+\! \frac{1}{s} \!\!\sum\limits_{m_r = 1}^{+ \infty} \!\!\frac{\mathcal{I}_{m_r} [\chi]}{1 \!-\! \big[ m_r / s \big]^{2}} \!\right] \, , \nonumber
\\
\!\!\!\! \displaystyle \mathcal{H} (s,\chi) = 2 \, (1 \!-\! s^2) \frac{e^{- \chi}}{\chi} \frac{1}{s} \!\!\sum\limits_{m_r = 1}^{+ \infty} \!\!\frac{m_{r} \, \mathcal{I}_{m_r} [\chi]}{1 \!-\! \big[ m_r / s \big]^{2}} \, , \label{definition_F_G}
\\
\!\!\!\! \displaystyle \mathcal{I} (s,\chi) = 2 \, (1 \!-\! s^2) \frac{e^{- \chi}}{\chi} \!\left[ \frac{1}{2} \frac{\mathcal{I}_{1} [ \chi ]}{s} \!+\! \frac{1}{s} \!\!\sum\limits_{m_r = 1}^{+ \infty} \!\!\frac{\mathcal{I}_{m_r + 1} [\chi]}{1 \!-\! \big[ m_r / s \big]^{2}} \!\right] \, , \nonumber
\end{numcases}
Moreover, we notice that we can use the simplification ${ \partial / \partial J_{\phi}  \!\!\left[ \kappa / \sigma_r^{2} \right] \sigma_r^{4} / \kappa^2 = - \partial / \partial J_{\phi}  \!\!\left[ \sigma_r^{2} / \kappa \right] }$, and that thanks to equation~\eqref{definition_intrinsic_frequencies}, one can also explicitly compute
\begin{equation}
\bigg| \frac{\mathrm{d} J_{\phi}}{\mathrm{d} R_g} \bigg|_{R_{0}} \!\!\!\!= \frac{R_g^{3} \kappa^{2}}{2 J_{\phi}} \bigg|_{R_{0}} \!\!\!\!= \frac{R_{0} \kappa^{2}}{2 \Omega} \, .
\label{calcul_dJphi_dRg}
\end{equation}
Finally, using the expression of the amplitude of the basis potentials from equation~\eqref{expression_amplitude_WKB}, we obtain a detailled expression of the matrix coefficients as
\begin{align}
& \widehat{\mathbf{M}}_{\left[  k_{\phi}^{p}, k_{r}^{p} , R_0 \right] , \left[ k_{\phi}^{q} , k_{r}^{q} , R_0 \right]} (\omega) = \delta_{k_{\phi}^{p}}^{k_{\phi}^{q}} \, \delta_{k_{r}^{p}}^{k_{r}^{q}} \, \frac{2 \pi G \Sigma |k_{r}| }{\kappa^{2} (1\!-\! s^{2} )} \, \times \nonumber
\\
\label{calcul_M_tight_IV} & \;\;\;\; \bigg\{\! \mathcal{F} (s , \chi) + k_{\phi}^{p} \frac{\sigma_{r}^{2}}{\kappa} \frac{\partial }{\partial J_{\phi}} \!\!\left[\! \ln \!\left(\! \frac{\Omega \Sigma}{\kappa \sigma_r^{2}} \!\right) \!\!\right] \mathcal{G} (s , \chi)
\\
& \;\;\;\; + k_{\phi}^{p} \frac{\partial }{\partial J_{\phi}} \!\!\left[ \frac{\sigma_{r}^{2}}{\kappa} \right] \!\left[ (1 \!-\! \chi) \mathcal{G} (s , \chi) \!+\! \mathcal{H} (s , \chi) \!+\! \chi \mathcal{I} (s , \chi) \right] \!\!\bigg\} \, , \nonumber
\end{align}
where one must remember that within the WKB approximation, the response matrix is diagonal. For a tepid disc, we may neglect some of the terms appearing in equation~\eqref{calcul_M_tight_IV}. A tepid disc corresponds to a disc where the orbits possess a small radial energy, so that all the orbits are close to circular orbits. It also implies that ${ \big| \partial F_{0} / \partial J_{r} \big| \gg \big| \partial F_{0} / \partial J_{\phi} \big| }$. For a Schwarzschild distribution function, the typical spread in $J_{r}$ is of the order of ${ \sigma_{r}^{2} / \kappa }$, so that we may consider equation~\eqref{calcul_M_tight_IV} as a limited development in ${ \sigma_r^{2} / \kappa }$ and ${ \partial / \partial J_{\phi} \!\!\left[ \sigma_{r}^{2} / \kappa \right] }$. Therefore, for a tepid disc the diagonal coefficients of the response matrix finally take the form given in equation~\eqref{calcul_M_tight_tepid}.

\section{Autocorrelation diagonalization}
\label{sec:autocorrelationdiagonalization}

Let us now show how the hypothesis of spatially quasi-stationarity of the external perturbations introduced in equation~\eqref{stationarity_translation_assumption_text} leads to a \textit{diagonalization} of the autocorrelation with respect to the radial frequencies $k_{r}$ as shown in equation~\eqref{diagonalization_autocorrelation}. In order to shorten the notations, we do not write anymore the dependence with respect to the azimuthal number $m_{\phi}$, and the exterior perturbation will be noted as ${ \psi = \psi^{e} }$. As a consequence, the assumption of temporal and quasi-spatial stationarity from equation~\eqref{stationarity_translation_assumption_text} takes the form
\begin{equation}
\langle \psi [R_{1} , t_{1}] \, \psi^{*} [R_{2} , t_{2}]\rangle = \mathcal{C} [t_{1} \!-\! t_{2} , R_{1} \!-\! R_{2} , (R_{1} \!+\! R_{2}) / 2] \, . 
\label{stationarity_translation_assumption_appendix}
\end{equation}
Equation~\eqref{simplification_diffusion_coefficients_II} for the diffusion coefficients requires us to study the term ${ \big< \widehat{\psi}_{k_{r}^{1}} [R_{g} , \omega_{1}] \, \widehat{\psi}^{\, *}_{k_{r}^{2}} [R_{g} , \omega_{2} ] \big> }$. Using the definition of the temporal Fourier transform from equation~\eqref{definition_Fourier_temporal} and the local radial Fourier transform from equation~\eqref{definition_local_Fourier}, we may rewrite it as
\begin{align}
\label{rewriting_terms_auto}  \big< \widehat{\psi}_{k_{r}^{1}} & [R_{g} , \omega_{1}] \, \widehat{\psi}^{\, *}_{k_{r}^{2}} [R_{g} , \omega_{2}] \big> =  
\\
& \frac{1}{4 \pi^{2}} \!\!\! \int \!\!\! \mathrm{d} t_{1}  \mathrm{d} t_{2} \mathrm{d} R_{1} \mathrm{d} R_{2} \, e^{i \omega_{1} t_{1}} e^{- i \omega_{2} t_{2}} \!\langle \psi [R_{1} , t_{1}]  \, \psi^{*} \![R_{2} , t_{2} ]\rangle  \, \times \nonumber
\\
& g [R_{g} \!-\! R_{1}] \, g [R_{g} \!-\! R_{2}] \, e^{-i (R_{1} - R_{g}) k_{r}^{1}} \, e^{i (R_{2} - R_{g})k_{r}^{2}} \, , \nonumber
\end{align}
where $g[R]$ is defined as
\begin{equation}
g [R] = \exp \left[ - R^{2} / (2 \sigma^{2})  \right] \, .
\label{definition_g_auto}
\end{equation}
We now use the assumption from equation~\eqref{stationarity_translation_assumption_appendix} relative to the radial dependences of the perturbation autocorrelation, and the change of variables
\begin{equation}
\begin{cases}
\begin{aligned}
\displaystyle & u_{t} = t_{1} \!+\! t_{2} \; &; \;\;\; & v_{t} = t_{1} \!-\! t_{2} \, ,
\\
\displaystyle & u_{r} = \tfrac{1}{2} (R_{1} \!+\! R_{2}) \; &; \;\;\; & v_{r} = R_{1} \!-\! R_{2} \, .
\end{aligned}
\end{cases}
\label{change_variables_auto}
\end{equation}
This transformation is of determinant $2$, so that equation~\eqref{rewriting_terms_auto} becomes
\begin{align}
\big< \widehat{\psi}_{k_{r}^{1}} & [R_{g} , \omega_{1}] \, \widehat{\psi}^{\, *}_{k_{r}^{2}} [R_{g} , \omega_{2}] \big> = \nonumber
\\
& \frac{1}{8 \pi^{2}} \!\! \int \!\! \mathrm{d} u_{t} \, \mathrm{d} v_{t} \, \mathrm{d} u_{r} \, \mathrm{d} v_{r} \, e^{ i \frac{\omega_{1} - \omega_{2}}{2} u_{t} } e^{i \frac{\omega_{1} + \omega_{2}}{2} v_{t}}  \, \times \nonumber
\\
\label{after_change_variables_auto} & e^{-i (k_{r}^{1} - k_{r}^{2}) u_{r}} e^{-i \frac{k_{r}^{1} + k_{r}^{2}}{2} v_{r}} e^{i R_{g} (k_{r}^{1} - k_{r}^{2})} \, \times
\\
& g [ R_{g} \!-\! u_{r} \!-\! v_{r}/2 ] \, g [ R_{g} \!-\! u_{r} \!+\! v_{r}/2 ] \, \mathcal{C} [v_{t} , v_{r}, u_{r}] \, . \nonumber
\end{align}
The integration on $u_{t}$ is straightforward and is equal to ${ 2 \pi \delta_{\rm D} ((\omega_{1} \!-\! \omega_{2})/2) }$. The integration on $v_{t}$ is then direct and gives ${ \widehat{\mathcal{C}} \, [\omega_{1} , v_{r} , u_{r}] }$. Finally, we note that the product of the two Gaussians in equation~\eqref{after_change_variables_auto} can be rewritten in order to disentangle the dependences on $u_{r}$ and $v_{r}$ to read
\begin{equation}
{ \!\! g[\!R_{g} \!\!-\! u_{r} \!\!-\!\! v_{r}\!/2] g[ \!R_{g} \!\!-\! u_{r} \!+\! v_{r}\!/2] \!=\! g[ \!\sqrt{\!2} (R_{g} \!\!-\! u_{r})] g[ \!v_{r} \!/\! \sqrt{\!2}] } \, ,
\label{product_Gaussians_auto}
\end{equation}
where the presence of $\sqrt{2}$ comes from the definition of the $g$ function introduced in equation~\eqref{definition_g_auto}. One can then  rewrite equation~\eqref{after_change_variables_auto} as
\begin{align}
\label{before_radial_integration_auto}  \big< \widehat{\psi}_{k_{r}^{1}} & [R_{g} , \omega_{1}] \, \widehat{\psi}^{\, *}_{k_{r}^{2}} [R_{g} , \omega_{2}] \big> =
\\
& \frac{1}{2 \pi} e^{i R_{g} (k_{r}^{1} - k_{r}^{2})} \delta_{\rm D} (\omega_{1} \!-\! \omega_{2}) \!\! \int \!\! \mathrm{d} v_{r} \, e^{- i \frac{k_{r}^{1} + k_{r}^{2}}{2} v_{r}} g[v_{r} \!/\! \sqrt{2}] \, \times \nonumber
\\
& \int \!\! \mathrm{d} u_{r} \, \widehat{\mathcal{C}} \, [\omega_{1} , v_{r} , u_{r}] \, g[\sqrt{2} (R_{g} \!-\! u_{r})] e^{- i (k_{r}^{1} - k_{r}^{2}) u_{r}} \, . \nonumber
\end{align}
As we have assumed that the function ${ u_{r} \mapsto \widehat{\mathcal{C}} \, [\omega_{1} , v_{r} , u_{r}] }$ is a slowly varying function, we may take it out of the integration on $u_{r}$ and evaluate it as ${ \widehat{\mathcal{C}} \, [\omega_{1} , v_{r} , R_{g}] }$. The remaining integration on $u_{r}$ can then be computed and reads
\begin{align}
\label{remaining_integration_ur} \int \!\! \mathrm{d} u_{r} \,  g[\sqrt{2} (R_{g} \!-\! u_{r})] & e^{- i (k_{r}^{1} - k_{r}^{2}) u_{r}} 
\\
& = \sqrt{\pi} \sigma e^{- i R_{g} (k_{r}^{1} - k_{r}^{2})} \exp \!\left[\! - \frac{(k_{r}^{1} \!-\! k_{r}^{2})^{2}}{4 / \sigma^{2}} \!\right] \,  \nonumber
\\
& = 2 \pi \, \delta_{\rm D} (k_{r}^{1} \!-\! k_{r}^{2}) \, e^{- i R_{g} (k_{r}^{1} - k_{r}^{2})} \, , \nonumber
\end{align}
where we replaced the Gaussian in ${ k_{r}^{1} \!-\! k_{r}^{2} }$ by a Dirac delta, while paying a careful attention to the correct normalization. As a consequence, equation~\eqref{before_radial_integration_auto} becomes
\begin{align}
\label{last_radial_integration_auto} \!\!\! \big< \widehat{\psi}_{k_{r}^{1}} [R_{g} , \omega_{1}] \, \widehat{\psi}^{\, *}_{k_{r}^{2}} [R_{g} ,  \omega_{2}] & \big> = \, \delta_{\rm D} (\omega_{1} \!-\! \omega_{2}) \, \delta_{\rm D} (k_{r}^{1} \!-\! k_{r}^{2}) \, \times
\\
&  \!\!\! \int \!\! \mathrm{d} v_{r} \, e^{- i k_{r}^{1} v_{r}} \widehat{\mathcal{C}} \, [\omega_{1} , v_{r} , R_{g}]  \,g [v_{r} \!/\! \sqrt{2} ]\, . \nonumber
\end{align}
Because of the definition from equation~\eqref{definition_local_Fourier}, the presence of the factor ${ 1\!/\!\sqrt{2} }$ corresponds to the change ${ \sigma \to \sqrt{2} \, \sigma }$ so that the remaining integral on $v_{r}$ may be interpreted as a local radial Fourier transform centered around the position ${ v_{r} = 0 }$. Therefore, we straightforwardly obtain the \textit{diagonalized} expression introduced in equation~\eqref{diagonalization_autocorrelation}.

\label{lastpage}
\end{document}